\documentclass[12pt,a4paper]{article}
\usepackage{jheppub}
\usepackage{amsmath,amssymb,euscript,array,mathrsfs,epsfig}
\usepackage[small]{caption}
\def\mg{{\mathfrak g}}

\newcommand{\EQ}[1]{\begin{equation}\begin{split} #1
\end{split}\end{equation}}
\def\ben
{\begin{equation}}
\def\een{\end{equation}}

\def\U{\text{U}}

    \let\L=\Lambda
   
 \let\W=\mu

\def\W={\cal W}
\def\L ={\cal L}

\def\be{\begin{equation}}
\def\ee{\end{equation}}
\def\ba{\begin{array}}
\def\ea{\end{array}}

\def\dalemb#1#2{{\vbox{\hrule height .#2pt
        \hbox{\vrule width.#2pt height#1pt \kern#1pt
                \vrule width.#2pt}
        \hrule height.#2pt}}}

\newcommand{\bea}{\begin{eqnarray}}
\newcommand{\eea}{\end{eqnarray}}

\newcommand{\mf}{\mathfrak}

\title{ Asymptotic symmetries and thermodynamics of higher spin black holes in AdS$_3$}
\author{Michael Ferlaino,} 
\author{Timothy J. Hollowood} 
\author{and S. Prem Kumar}
\affiliation{Department of Physics,\\Swansea University, \\ 
Singleton Park,\\ Swansea, SA2 8PP, U.K.}
\emailAdd{pymf@swansea.ac.uk} 
\emailAdd{t.hollowood@swansea.ac.uk}
\emailAdd{$\qquad\quad\,\,$s.p.kumar@swansea.ac.uk}

\abstract{We study black holes  carrying higher spin charge  
in AdS$_3$ within the framework of  ${\rm SL}(N,\mathbb R) \times {\rm SL}(N,{\mathbb R})$ Chern-Simons theory. Focussing attention on the $N=4$ case, we explicitly analyze the asymptotic symmetry algebra of black hole solutions with a chemical potential for spin-four charge. We demonstrate that the background  describes an RG flow between an IR fixed point  with ${\cal W}_4$ symmetry and a UV fixed point with ${\cal W}$-symmetry associated to a non-principal embedding of $\mf{sl}(2)$ in $\mf{sl}(4)$. Matching Chern-Simons equations with Ward identities of the deformed CFT, we show that the UV stress tensor is twisted by a certain $\U(1)$ current, and the flow is triggered by an operator with dimension $4/3$ at the UV fixed point. We find independent confirmation of this picture via a consistent formulation of thermodynamics with respect to this UV fixed point.
We further analyze the thermodynamics of multiple branches of black hole solutions for $N=4,5$ and find that the BTZ-branch, dominant at low temperatures, ceases to exist at higher temperatures following a merger with a thermodynamically unstable branch. We also point out an interesting connection between the RG flows and generalized KdV hierarchies.}
\begin{document}
\maketitle
\flushbottom
\section{Introduction}
Theories of gravity with interacting higher spin gauge fields have come to play a central role  within the general theme of the AdS/CFT correspondence
\cite{maldacena, witten}.  Large-$N$ gauge theories in the free limit, dual to tensionless limits of string theory, require the presence of infinite towers of higher spin gauge fields in the dual string theory \cite{Konstein:2000bi, HaggiMani:2000ru, Sundborg:2000wp, Mikhailov:2002bp, Sezgin:2002rt}. The most concrete realizations of similar dualities involving higher spin gauge theory have been shown to arise in the context  of large-$N$ vector models
in three \cite{Klebanov:2002ja,Giombi:2009wh,Giombi:2010vg,Chang:2012kt} and two dimensions \cite{gg, Gaberdiel:2011zw, Gaberdiel:2011wb, Chang:2011mz, Gaberdiel:2012ku, Gaberdiel:2012uj}. Within the setting of AdS$_3$/CFT$_2$ duality, which is the focus of the present paper, the complexity of higher spin theories can be reduced as it becomes possible to truncate to a finite set of higher spin fields and formulate the system as ${\rm SL}(N,\mathbb R) \times{\rm SL}(N, \mathbb R)$ Chern-Simons theory \cite{blencowe, Bergshoeff:1989ns,Bordemann:1989zi,Vasiliev:1989qh,Vasiliev:1989re,Fradkin:1990qk,Henneaux:2010xg,campoleoni}. These theories provide us not only with an opportunity to explore the holographic duality within a novel, tractable framework, but also allow us to study properties of black hole like objects in theories where diffeomorphism invariance is enlarged to a higher spin gauge symmetry. 
 
Black hole configurations carrying spin-three charge in ${\rm SL}(3,\mathbb R) \times {\rm SL}(3, \mathbb R)$ Chern-Simons theory were constructed in \cite{gutkraus}, and shown to have several remarkable features. Notably, the black hole horizon ceases to be a gauge-invariant notion \cite{ammon}, and whether a configuration is a black hole or not is determined by the holonomy of the Chern-Simons connection around the Euclidean time circle.
Indeed, fixing this holonomy guarantees that the first law of thermodynamics holds for the higher spin black holes \cite{deboer,conical2}.

Another notable feature of the higher spin black holes as constructed in \cite{gutkraus} is that a non-vanishing higher spin charge is obtained only when a chemical potential for the corresponding higher spin current is also turned on. In the language of the dual CFT$_2$, a chemical potential for any current with spin $s>2$ can be viewed as an {\em irrelevant} deformation by an operator of dimension $s$. On general grounds such a deformation 
would spoil the ultraviolet (UV) behaviour of any  CFT$_2$, and the associated grand canonical partition function can be viewed at best as an asymptotic expansion in powers of the higher spin chemical potential. It is surprising, therefore, that in the dual gravity picture of \cite{gutkraus, ammon, review}, while the higher spin chemical potential does alter the asymptotic AdS$_3$ (UV) geometry, the alteration is of a special nature: it corresponds to a flow to a new AdS$_3$ with a different radius and a different asymptotic symmetry algebra. Specifically, in the 
${\rm SL}(3, \mathbb R)\times {\rm SL}(3, \mathbb R)$ Chern-Simons theory, the spin-three chemical potential induces a flow between a theory with ${\cal W}_3$ symmetry and a different UV theory with ${\cal W}_3^{(2)}$ symmetry (the Polyakov-Bershadsky algebra \cite{Polyakov:1989dm, Bershadsky:1990bg}). These two symmetry algebras correspond to the two different embeddings of $\mf{sl}(2)$ in $\mf{sl}(3)$, the so-called principal and non-principal (or diagonal) embeddings, respectively.

The picture above is expected to naturally generalize for any (integer) $N$. However, the specifics of such generalizations are yet to be  explored completely. It is not {\em a priori} obvious whether the new (UV) AdS$_3$ asymptotics, induced by a higher spin chemical potential, will always have a ${\cal W}$-algebra symmetry associated to a non-principally embedded ${\rm SL(2,\mathbb R)}$ in ${\rm SL}(N,\mathbb R)$. For instance, as pointed out in \cite{tan}  the naive central 
charge corresponding to a specific non-principal embedding is at odds with the value inferred from the AdS-radius of the asymptotic geometry, for generic $N$. Understanding the nature and symmetries of these UV fixed points is important if the grand canonical ensemble (on the gravity side) is to be taken seriously for generic finite values of the higher spin chemical potential. Eventually we would also want to make sense of the asymptotics of analogous black hole solutions within the ${\rm hs}[\lambda]$ Vasiliev theory \cite{krausperl, Prokushkin:1998bq} dual to ${\cal W}_N$ minimal models in the 't~Hooft large-$N$ limit \cite{gg,Gaberdiel:2012ku}.

In this paper we investigate the theory with $N=4$, deformed by a chemical potential for spin-4 charge. 
We perform an explicit analysis of the asymptotic symmetry algebra (classical Drinfeld-Sokolov reduction) for the resulting Chern-Simons connections, along the lines of \cite{campoleoni}. We find that the black hole solution\footnote{The spin-4 black hole solution we study was first obtained in \cite{tan}.}
indeed describes a flow between two conformal fixed points, with two different ${\cal W}$-algebras. While the original theory has ${\cal W}_4$ symmetry, the new UV fixed point exhibits a 
${\cal W}$-symmetry associated to the $(2,1,1)$ 
embedding\footnote{Embeddings of $\mf{sl}(2)$ in $\mf{sl}(N)$ are classified by integer partitions of $N$. }
 of $\mf{sl}(2)$ in $\mf{sl}(4)$. 
 
Importantly, we find that conformal transformations at the UV fixed point are generated by a {\em twisted\/} stress tensor, where the twisting/improvement is by an abelian current in the asymptotic symmetry algebra. 
The stress tensor improvement is inferred by requiring  the Chern-Simons equations of motion to match with Ward identities in the UV fixed point theory. Precisely the same twisting is also necessary in order to explain why the ratio of central charges of the UV and IR ${\cal W}$-algebras is determined by the ratio of the corresponding AdS radii. An immediate consequence of this is that the black hole solution can be viewed as a flow induced by a relevant deformation of the UV fixed point, via a chemical potential for a conserved current (with scaling dimension $4/3$). Using the approach of \cite{bct, David:2012iu}, we further show that formulating  thermodynamics with respect to the UV fixed point independently confirms the results obtained from the analysis of the asymptotic symmetry algebra and Ward identities.

The existence of multiple branches of solutions to the holonomy conditions which determine a black hole background implies a rich thermodynamic phase structure which was explored in \cite{David:2012iu} within the ${\rm SL}(3,{\mathbb R})$ Chern-Simons theory (see also \cite{Chen:2012pc,Chen:2012ba}). Extending this to ${\rm SL}(4,\mathbb R)$ and ${\rm SL}(5,\mathbb R)$ theories, we find that there is a general pattern. The BTZ-branch of solutions\footnote{
These reduce to the BTZ black hole for zero higher spin chemical potential.} dominate the ensemble for low temperatures (equivalently, low chemical potential for fixed temperature), and eventually merge with a thermodynamically unstable branch and cease to exist beyond a critical temperature. In addition, the number of (real) branches increases rapidly with $N$.

Our analysis confirms the expectation that higher spin black hole solutions in ${\rm SL}(N,\mathbb R)$ Chern-Simons theories are generically embedded within RG flows between an IR CFT with ${\cal W}_N$ symmetry and a UV CFT with a non-principal ${\cal W}$-symmetry. 
In general, the stress tensor of the UV theory is twisted by an abelian current in such a way that the resulting ratio of UV and IR central charges agrees with the ratio of the UV and IR AdS-radii. 
In conjunction with the fact that the number of non-principal embeddings grows rapidly with $N$ (as $e^{\sqrt N}$), this makes non-principal embeddings important to investigate. It has been noted that in the semiclassical limit (fixed $N$ and large Chern-Simons level), non-principal embeddings can lead to non-unitary theories with negative norm states \cite{Castro:2012bc}. However, it is also possible to arrive at semiclassical limits (large central charges) maintaining unitarity for certain types of non-principal embeddings \cite{Afshar:2012hc}.

The paper is organized as follows: In Section 2 we lay out our notation and conventions, and review some simple features of flows induced by higher spin chemical potentials. The goal of Section 3 is to obtain the asymptotic symmetry algebra for the spin-4 black hole in ${\rm SL}(4,\mathbb R)$ Chern-Simons theory. This includes matching of Chern-Simons equations of motion to the Ward identities of the deformed UV CFT. In Section 4, we set up and analyze the thermodynamics of multiple black hole branches in ${\rm SL}(4)$ and 
${\rm SL}(5)$ Chern-Simons theory. Technical details of the computations are relegated to Appendices A, B, C and D.

\section{SL$(\boldsymbol N,{\mathbb R})\times\text{SL}(\boldsymbol N,{\mathbb R})$ Chern-Simons Connections}

It is a remarkable fact that Einstein gravity in three dimensions, with a negative cosmological constant, can be reformulated as ${\rm SL}(2, \mathbb R)\times {\rm SL}(2, \mathbb R)$ Chern-Simons theory on a three-manifold ${\cal M}$ \cite{Achucarro:1987vz, Witten:1988hc}. It has now been realized that, more generally, ${\rm SL}(N, \mathbb R)\times {\rm SL}(N, \mathbb R)$ Cherns-Simons theory on ${\cal M}$ is an interacting theory of gravity and a tower of higher spin fields on ${\rm AdS}_3$ \cite{campoleoni, Henneaux:2010xg}. 

Introducing the coordinates $(\rho,z,\bar z)$ on ${\cal M}$, asymptotically AdS$_3$ backgrounds in ordinary gravity correspond to flat connections in ${\rm SL}(2)\times{\rm SL}(2)$ Chern-Simons theory of the form
\be
A\,=\,L_0\,d\rho + \left(L_1\,e^\rho +\cdots \right)\, dz\,,\qquad
\bar A\,=\, - L_0\,d\rho + \left(- L_{-1}\, e^\rho +\cdots \right)\,d\bar z \,.
\label{asymp}
\ee
where the ellipsis indicate subleading terms in the large $\rho$ limit. 
We denote the $\mf {sl}(2,\mathbb R)$ generators as $\{L_0, L_{\pm1}\}$, satisfying,
\be
[L_1, L_{-1}]\,=\,2L_0\,,\qquad[L_{\pm 1}, L_0]\,=\,\pm\,L_{\pm 1}.
\ee
The complex coordinates $(z, \bar z)$ are related to the boundary Euclidean time $t$ and spatial coordinate $\phi\simeq \phi +2\pi$ as
\be
z\,\equiv\,it+\phi\,,\qquad\qquad\bar z\,\equiv\,it-\phi\,.
\ee
There are several inequivalent ways of embedding $\mf{sl}(2)$ in $\mf{sl}(N)$, and each embedding is naturally associated to a distinct AdS$_3$ asymptotics, possibly dual to a distinct conformal fixed point field theory. Of these embeddings, the so-called principal embedding is special and the resulting theory can be viewed as Einstein gravity on ${\rm AdS}_3$ coupled to higher spin fields with spins $s =3,4,\ldots,N$. The principal embedding  results upon choosing the generators $L_0, L_{\pm1}$ appearing in eq.~\eqref{asymp} to form the $N$-dimensional irreducible representation of ${\mf {sl}}(2)$.

The work of \cite{gutkraus} showed that even if one were to focus attention on the principal embedding to formulate a higher spin theory of gravity on AdS$_3$, with an asymptotic ${\cal W}_N\times {\cal W}_N$ symmetry algebra, one may be forced to also consider theories arising from non-principal embeddings of ${\mf {sl}}(2)$. At least, this is the case for the ${\rm SL}(3,{\mathbb R}) \times {\rm SL}(3,{\mathbb R})$ Chern-Simons theory. In particular, black hole solutions carrying a spin-3 charge necessarily modify the AdS$_3$ asymptotics by inducing a flow that connects to a new UV fixed point whose symmetry algebra, ${\cal W}_3^{(2)}\times {\cal W}_3^{(2)}$, is associated to a non-principal embedding of ${\mf {sl}}(2)$ in ${\mf {sl}}(3)$.

Two questions naturally follow from the considerations above: 
(i) whether the construction of higher spin black holes can be generalized to ${\rm SL}(N, {\mathbb R})$ Chern-Simons theory, and, (ii) whether every such higher spin generalization involves an RG flow between two CFTs with different ${\cal W}$-algebras\footnote{For a discussion of RG flows in higher spin supegravity on ${\rm  AdS}_3$, see \cite{Peng:2012ae}}. Attempts to answer similar questions have already been made in \cite{krausperl} and \cite{tan}. Our motivation is to focus on these issues carefully, and then to further explore the thermodynamic phase structure of such higher spin solutions, and possibly generalize the results of \cite{David:2012iu}.

\subsection{Higher spin chemical potentials and zero temperature flows}

Following the conventions of \cite{conical} for the principal embedding of ${\mf{sl}}(2)$, we label the generators of $\mf{sl}(N)$ as $\{L_0,L_{\pm1}\}$ and $\{W_m^{(s)}\}$ with $m=-(s-1),\ldots (s-1)$, for all $s=3,4,\ldots N$:
\be
[L_i, L_j]\,=\, (i-j) L_{i+j}\,,\qquad\qquad [L_i,W_m^{(s)}]\,=\,(i(s-1)-m)\,W_{i+m}^{(s)}\,.
\ee
The $\{L_i\}$ constitute the $N$-dimensional representation of $\mf{sl}(2)$, and the generators $W_m^{(s)}$ have weight $-m$ with respect to $L_0$. Matrix representations for all the generators can be found as outlined in Appendix \ref{conventions}.

The asymptotically AdS Chern-Simons connections $(A, \bar A)$ can be re-expressed in terms of $\rho$-independent flat connections $(a,\bar a)$, where the dependence on the radial coordinate is obtained via a gauge transformation generated by $L_0$
\bea
A \,=\, b^{-1}\,a b + b^{-1} db\,,\qquad
\bar A \,=\, b\,\bar a\, b^{-1} + b\,db^{-1}\,,\qquad
b= e^{L_0\rho}\,.
\eea
A chemical potential $\mu_s$ for spin-$s$ charge can be viewed as a deformation of the ${\cal W}_N$ CFT by the spin-$s$ current operator $\mf {W}_s(z)$,
\be
I_{\rm CFT}\,\to\, I_{\rm CFT}\, +\,\int d^2z\left(\mu_s\,{\mf W}_s\, +\, \bar\mu_s \,\overline{\mf W}_s\right),
\ee
where we will assume $\bar \mu_s=\mu_s$.
The constant Chern-Simons connections,
\be
a\,\equiv\,a_z\,dz\,+\,a_{\bar z}\,d\bar z\,,\qquad\qquad\qquad
\bar a\,\equiv\,\bar a_z\,dz\,+\,\bar a_{\bar z}\,d\bar z\,,
\ee
are then changed accordingly to reflect the deformation by the spin-$s$ current. In particular, the equations of motion (flatness conditions) obeyed by the connections must reproduce the Ward identities for various currents in the deformed CFT. This procedure has been explicitly demonstrated for ${\rm SL}(3,\mathbb R)$ in \cite{gutkraus,ammon, review} and for the ${\rm SL}(4,\mathbb R)$ Chern-Simons connections in \cite{tan}.

The upshot of these results is that in the presence of chemical potentials for higher spin charges, the (zero temperature) Chern-Simons connections acquire both holomorphic and anti-holomorphic components,
\bea
&&a_z\,=\,L_1\,,\qquad \qquad\quad a_{\bar z}\,=\,\sum_{s=3}^N \mu_s\, W_{s-1}^{(s)}\,,
\label{connection1}
\\\nonumber
&&\bar a_{\bar z}\,=- \,L_{-1}\,,\qquad\qquad \bar a_z\,=\,
\sum_{s=3}^N (-1)^{(s-1)}\,\mu_s\, W_{-s+1}^{(s)}\,,\\\nonumber\\\nonumber
&&[a_z,a_{\bar z}]\,=\,[\bar a_z,
\bar a_{\bar z}]\,=\,0\,.
\eea
By construction, these are constant, flat connections and encode the RG flows induced by deformations of the conformal field theory by $(N-2)$ higher spin conserved currents. Since a spin-$s$ current, $\mf W_s$, is a local operator of dimension $s \geq 3$, these are irrelevant deformations of the CFT, a fact reflected in the $\rho$-dependent gauge fields $(A,\bar A)$, 
\bea
&&A\,=\, e^{\rho}\,L_1\, dz + \,\left(\sum_{s=3}^N \mu_s\, W_{s-1}^{(s)}\, e^{(s-1)\rho}\right)\,d\bar z\,+ L_0\,d\rho\,,
\label{connections}\\\nonumber
&&\bar A\,=\,- e^{\rho}\,L_{-1}\,d\bar z +\,
\left(\sum_{s=3}^N (-1)^{(s-1)}\,\mu_s\, W_{-s+1}^{(s)} \, e^{(s-1)\rho}\right)\,dz\,- L_0\,d\rho\,.
\eea
Higher spin chemical potentials modify the asymptotics of the original AdS$_3$ spacetime, as would be expected for deformations of the CFT by irrelevant operators.
The spacetime metric is obtained from the gauge connections in a standard way,
\be
ds^2=\frac{1}{4\,\epsilon_N}\,{\rm Tr}\left[(A-\bar A)^2\right]\,,\qquad {\,\epsilon_N}\,=\,{\rm Tr}\,L_0^2\,= \tfrac{1}{12}\,N(N^2-1).
\ee
With the higher-spin chemical potentials switched on, at zero temperature, we find,
\be
ds^2\,=\,d\rho^2 +(d\phi^2-dt^2)\,\left(e^{2\rho}+
\sum_{s=3}^N\,\,e^{2(s-1)\rho}\,\,\mu_s^2\,\frac{t^{(s)}_{s-1}}{4\epsilon_N}\right)\,.\label{flowmetric}
\ee
The numerical coefficients $t^{s}_{(s-1)}$ are not important at this stage, and can be read off from the expression for the Cartan-Killing form for ${\rm SL}(N, {\mathbb R})$ given in Appendix A of \cite{conical}.

The metric \eqref{flowmetric} represents a flow between two different AdS geometries. Specifically, for non-vanishing 
$\mu_N$, it interpolates between an ${\rm AdS}_3$ with unit radius ( $e^\rho\ll 1$) and one with radius $\tfrac{1}{N-1}$ ($e^\rho \gg 1$). Along this flow the system passes by other putative fixed points dual to ${\rm AdS}_3$ geometries with radii $\tfrac{1}{s-1}$, for every non-zero $\mu_s$. It is not {\em a priori} obvious what the properties of CFTs dual to these new AdS$_3$ geometries are.
It has been shown for the specific case of ${\rm SL}(3,\mathbb R)$, that deformation by the spin-3 current leads to a UV conformal fixed point with ${\cal W}_3^{(2)}$ symmetry \cite{gutkraus, 
ammon, David:2012iu}.

Applying simple scaling arguments to the metric \eqref{connections} for large $\rho$,  we learn that the flow is generated by deforming the putative UV fixed point (dual to AdS$_3$ with radius $\tfrac{1}{N-1}$) by a tower of relevant operators with dimensions 
\be
\Delta_{\tiny{\rm UV}}\,=\,\frac{N+s-3}{N-1},\qquad s=3,4,\ldots, N\,.
\ee
Interestingly, in the large-$N$ limit these operators form a continuum with $1< \Delta_{\rm UV} < 2$.

It would be natural to assume that  the new putative fixed points in the UV should be related to non-principal embeddings of ${\mf {sl}}(2)$ in ${\mf{sl}}(N)$, generalizing the picture for $N=3$. However, certain doubts about the possibility of such an interpretation have been raised (see e.g.\cite{tan}). Discussion of this point will be crucial to our analysis.

\subsection{Non-principal embeddings}\label{tt}

The asymptotic UV form of the connections
\eqref{connections} with only a single chemical potential $\mu_s$, suggests a potential relation to a non-principally embedded ${\mf {sl}}(2)$ in $\mf{sl}(N)$ with generators $\{\hat L_{\pm1},\hat L_0\}$.
The step generators are related to the two elements $W^{(s)}_{\pm(s-1)}$ which take the form 
 \cite{conical} (a `$*$' indicates a non-zero matrix entry),
\bea
W_{-s+1}^{(s)}\,=\,\left(\begin{array}{ccccc}
~0~ & \cdots & * &\cdots&0  \\
0&0&\cdots& * &0\\
0&0 & 0 &\cdots &*\\
\vdots&\vdots&\vdots&\ddots& \vdots\\
0&0 & 0 & \cdots&0
\end{array}
\right)\,,\qquad
W_{s-1}^{(s)}\,=\,\left(\begin{array}{ccccc}
0 & 0 & ~0~ &\cdots&0  \\
\cdots&0&0&\cdots&0\\
*&\cdots & 0 &\cdots &0\\
\vdots&*&\vdots&\ddots& \vdots\\
0&0 & * & \cdots&0
\end{array}
\right)\,.
\eea
The relation with the $\mf{sl}(2)$ embedding is that, up to an overall conjugation,
\EQ{
\hat L_1=W^{(s)}_{s-1}\ ,\qquad \hat L_{-1}=UW^{(s)}_{-s+1}U^{-1}\ ,
}
where $U$ is a particular element of the Cartan subgroup $U=\text{diag}(\lambda_1,\ldots,\lambda_N)$ needed to ensure that the commutator of $\hat L_1$ with $\hat L_{-1}$ has the correct form.

In general different embeddings of $\mf{sl}(2)$ in $\mf{sl}(N)$ are 
classified by the partitions of $N$ \cite{dynkin, deBoer:1993iz, conical, Castro:2012bc}. While the principal embedding corresponds to the partition $(N)$, the more general embeddings with $\hat L_{1}=W^{(s)}_{s-1}$ correspond to other partitions of the form $(p_1^{n_1},p_2^{n_2})$. 
Below, we will  restrict attention to the case with $s=N$, i.e. when the CFT is deformed by a chemical potential for the highest spin current $(s=N)$. This has the advantage of being technically simpler, but nevertheless illustrates the more general features that we wish to understand. In this case, the relation with the $\mf{sl}(2)$ has $\hat L_{\pm1}
=W^{(N)}_{\pm(N-1)}$. The partition in this case is $(2,1^{N-1})$ and the branching of the adjoint of $\mf{sl}(N)$  in terms of a direct sum of irreducible representations of the $\mf{sl}(2)$ is
\be
\underline{\bf N^2-1}^{(2,1^{N-1})}\,\to\, \underline{\bf 3} + 2(N-2)\cdot\underline{\bf 2} + (N-2)^2\cdot\underline{\bf 1}
\ee 
The embedding defines the semi-simple decomposition of the algebra as
\be
\mf{sl}(N)\,\to\,\mf{sl}(2)\,\oplus\,\mf{sl}(N-2)\,\oplus
\,\mf{u}(1)\,,
\ee
and thus the asymptotic symmetries should include $\mf{sl}(N-2)$ and $\mf u(1)$ affine algebras.

However, the relation of the UV asymptotics of \eqref{connections} to such non-principal embeddings is likely to be subtle, as we now clarify.
With only a spin-$N$ chemical potential, $\mu$, the UV asymptotics is controlled by
\bea
a_{\bar z}\,\sim\,\left(\begin{array}{cccc}
0 & \ldots & 0 &0  \\
0&0&\ldots& 0 \\
0&0&0& \ldots\\
\mu&0 & 0 & 0
\end{array}
\right)\,,\qquad
\bar a_z\,\sim\,\left(\begin{array}{cccc}
0 & 0 & 0 &\mu  \\
\ldots&0&0&0\\
0&\ldots & 0 &0 \\
0&0&\ldots& 0
\end{array}
\right)\,.
\label{uv1}
\eea
These are proportional to the generators $W^{(N)}_{N-1}$ and $W^{(N)}_{-N+1}$, respectively.
For this case, the $\rho$-dependent gauge connection has the form (after a coordinate rescaling),
\bea
&&A\,=\, \left(W_{N-1}^{(N)}\,e^{\rho} +\ldots\right)\,dz + \tfrac{1}{N-1}L_0\, d\rho\,,\label{uv2}\\\nonumber\\\nonumber
&&\bar A\,=\, \left(- W_{-N+1}^{(N)}\,e^{\rho} +\ldots\right)\,d\bar z - \tfrac{1}{N-1}L_0\, d\rho\,.
\eea
Here we encounter the crucial point: the connections involve $L_0$, which differs from the generator of the two dimensional representation of $\mf{sl}(2)$, namely $\hat L_0$:
\be
\tfrac{1}{N-1}\,L_0\,=\,\tfrac{1}{2}\,{\rm diag}(1\,,\,\tfrac{N-3}{N-1}\,, \ldots, -\tfrac{N-3}{N-1}\,,\,-1)\,,\qquad
\hat L_0\,=\,{\rm diag}(\tfrac{1}{2},0,0,\ldots,\tfrac{1}{2})\,.
\ee
For this reason it is not {\em a priori} clear what the symmetries associated to this UV AdS$_3$ should be. Even if there is a UV conformal fixed point theory, the ratio of UV and IR central charges cannot simply be given by the ratio of the quadratic Casimirs of the representations, namely
${\rm Tr}(\hat L_0^2)/{\rm Tr}(L_0^2) = 6/(N(N^2-1))$. We conjecture that the ratio of central charges is actually controlled by the square of the AdS radii as,
\be
c_{\rm UV}\,=\,\frac{1}{(N-1)^2}\,c_{\rm IR}\,.
\label{ratioc}
\ee
We will see explicitly by computing the UV Ward identities for $N=4$ that this expectation is borne out. For the specific case of $N=3$, both expressions yield the same result, $c_{\rm UV}\,=\,c_{\rm IR}/4$. 
The above observations indicate that the asymptotic symmetry algebra implied by the Chern-Simons connection \eqref{uv2} needs to be carefully examined. 
In fact $L_0$ differs from the two dimensional
generator $\hat L_0$ by a re-scaling and some $\U(1)$ element, say $J_0$,
\be
\frac{1}{N-1}\,L_0\,=\,\hat L_0 + J_0\,.
\ee
Although this is puzzling at first sight, such a shift can be easily accounted for in the CFT 
by an ``improved" stress tensor,
\be
T \to \widetilde T \,=\,T+ \partial {\cal J}_0\,,
\ee
where ${\cal J}_0$ is an abelian current with zero mode $J_0$.
In such a situation, the conformal transformations are actually generated by a twisted stress tensor $\widetilde T$ which has a different central charge as compared to the ``canonical" stress tensor. We will show below, by explicit evaluation of the asymptotic symmetry algebra for the $(2,1,1)$ embedding in $\mf {sl}(4)$, the emergence of precisely such a twisted stress tensor in the ${\cal W}$-algebra of the UV CFT.
In the general case, it is worth noting that the element $L_0$ provides a consistent gradation of the whole algebra \EQ{
[L_0,W^{(s)}_{m}]\,=\,-m W^{(s)}_{m}
}
and for the non-principal embedding with $\hat L_1=W^{(s)}_{s-1}$ one has
\EQ{
\frac1{s-1}L_0=\hat L_0+J_0^{(s)}\ ,
}
where $J_0^{(s)}$ is some Cartan element and one expects that the stress tensor of the UV fixed point is similarly improved with $c_\text{UV}=c_\text{IR}/(s-1)^2$.

\section{Asymptotic UV Symmetry for SL$(\boldsymbol 4,\mathbb R)$ Black Hole}

In this Section we focus on the spin-4 black hole solution in ${\rm SL}(4,{\mathbb R})$ Chern-Simons theory. This solution was first obtained in \cite{tan}, and many of its features pointed out. Our motivation for revisiting this solution is two-fold. The black hole background provides a flat connection  whose asymptotics (IR and UV) have the form dictated by eqs.~\eqref{connections} and \eqref{uv1}. The spin-4 black hole is described by the constant connection in the highest weight representation,
\bea
&& a\,=\,a_z\,dz\,+\,a_{\bar z}\, d{\bar z}
\label{sl4}\\\nonumber\\\nonumber
&& a_z\,=\, \left(L_1\, -\, \tfrac{\pi }{5k}\,{\cal L}\,L_{-1}\,-\tfrac{\pi}{12k}\,Q_3\, W_{-2}^{(3)}\, +\,\tfrac{\pi}{18 k}\, {Q_4}\, W_{-3}^{(4)}\right)\\\nonumber\\\nonumber
&&a_{\bar z}\,=\,\mu\,\left(W_3^{(4)} \,-\, \tfrac{3\pi }{5k}\,{\cal L}\,W_{1}^{(4)}\,+\,\tfrac{\pi }{2k}\,{ Q_3}\,W_0^{(3)}\,+\, 
\tfrac{3\pi}{5 k}\,{Q_4}\, L_{-1}
\,+\,\left( \tfrac{3\pi^2}{25 k^2}\,{\cal L}^2\,
+\, \tfrac{\pi }{6 k}\,{Q_4}\right)\, W_{-1}^{(4)}\right.\\\nonumber\\\nonumber
&&\left.\qquad\,-\,
\tfrac{2\pi^2}{25 k^2}\,{\cal L}{Q_3}\,W_{-2}^{(3)}
\,-\,\left(\tfrac{\pi^2}{12k^2}\,{Q_3}^2\, +\,\tfrac{\pi^3}{125k^3}\,{\cal L}^3\,
-\,\tfrac{11\pi^2}{450k^2}\,{\cal L}{Q_4}\right)\,W_{-3}^{(4)}\right)\,,
\eea
with a similar expression for the barred connection. Here $Q_3, Q_4$ are the spin-3 and spin-4 charges respectively, and 
$\mu$ is a chemical potential conjugate to the spin-4 charge.
The flatness of the connection, $[a_z,a_{\bar z}]=0$, implies that $a_{\bar z}$ may be expressed as a traceless function of $a_z$, and for the spin-4 black hole we have
\be
a_{\bar z}\,=\,-\tfrac{41\pi}{25k}\,{\cal L}\,\mu\,a_z \,+\, \mu\,\left(a_z^3-\frac{1}{4}{\rm Tr}\,a_z^3\right)\,.
\ee
The thermodynamic charges ${{\cal L}, Q_3}$ and $Q_4$ are then related to  traces of powers of the holomorphic component of the connection $A$,
\bea
&& 2\pi\,{\cal L}\,=\,\frac{k}{2}\,{\rm Tr}\,(a_z)^2\,,\\\nonumber
&&-2\pi\,Q_3\,=\,\frac{k}{3}\,{\rm Tr}\,(a_z)^3\,,\\\nonumber
&&-2\pi\,Q_4\,+\,\tfrac{41\pi^2}{25k}{\cal L}^2\,=\,\,\frac{k}{4}\,{\rm Tr}\,(a_z)^4\,.
\eea
As pointed out in \cite{bct}, the relation between thermodynamic charges and the traces of the connection is quite general and one may define the charges for a given 
${\rm SL}(N,\mathbb R)$ gauge field in terms of the traces $\,{\rm Tr}\,(a_z^p)\,$, $p=2,3,\ldots,N$.

The values of the thermodynamic charges as a function of temperature and chemical potential are fixed by the smoothness condition on the holonomy of the gauge field around the thermal circle \cite{gutkraus, review, conical2}. We will return to this when we discuss the thermodynamics of higher spin black holes. For now, our main focus is establishing the ${\cal W}$-algebra of the UV conformal fixed point associated to the flat $\mf{sl}(4)$ connection controlling the asymptotics of the black hole solution.

\subsection{UV Ward identities}

In the original works \cite{gutkraus,ammon} the general form of the Chern-Simons connection describing a higher spin black hole was used to obtain the Ward identities of the CFT deformed by a chemical potential for a higher spin current. The basic strategy is to consider the most general ansatz for the gauge connection in the highest weight representation, allowing all the parameters (charges and potentials) to acquire dependence on the boundary coordinates $(z,\bar z)$. The equations of motion satisfied by these, following from the flatness of the connection, can then be matched onto the Ward identities for the currents of the appropriate ${\cal W}$-algebra in the CFT deformed by a higher spin chemical potential.

In order to perform this analysis for the UV fixed point implied by the gauge connection, after a co-ordinate rescaling (a constant gauge transformation) we write the ${\rm SL}(4,\mathbb R)$ gauge field \eqref{sl4} in the general form,
\bea
&&a\,=\,\Big(\tfrac{1}{6}\,W_{3}^{(4)}
\,+\sum_{\ell=-3}^2\,w_{\ell}\,W_{\ell}^{(4)}
\,+\,\sum_{\ell =-2}^2 v_\ell\,W_{\ell}^{(3)} + 
u_{0}\,L_0 + u_{-1}\, L_{-1}\Big)
\,d\bar z\label{uvsl4}\\\nonumber\\\nonumber
&&\qquad+\,\Big(\lambda \,L_1\,+\, \lambda_{-1}\, L_{-1}\,+ \,\lambda_{-2}\,W_{-2}^{(3)}\,+\,\lambda_{-3}\,W_{-3}^{(4)}\Big)\,dz\,.
\eea
The analysis for the barred gauge field $\bar a$ proceeds identically, so we restrict attention to the unbarred sector.
The constant gauge transformation in question is generated by $L_0$, so that 
$a \to e^{-\Lambda L_0}\,a\, e^{\Lambda L_0}$ with $e^\Lambda\,=\,(6\mu)^{-1/3}$ and
\be
\lambda\,=\, (6\mu)^{-1/3}\,.
\ee
Rewriting the connection \eqref{sl4} in this form, allows to switch our perspective from the IR to the UV. In particular, for large radial coordinate $\rho$, the $a_{\bar z}$ component provides an AdS$_3$ ``background" of radius $1/3$, whilst $a_z$ acts as a relevant ``deformation". The parameters appearing in the gauge field are related simply to those in \eqref{sl4} via rescalings by powers of $\mu$.

If we allow all deformation parameters in \eqref{uvsl4} to depend on the boundary co-ordinates $(z,\bar z)$, the flatness conditions
\be
da\,+\,a\wedge a\,=\,0\,,
\ee
are satisfied if certain constraints are imposed on the deforming parameters,
\bea
&&\bar\partial\lambda\,=\,0\,,\qquad w_2\,=\,v_2\,=v_1\,=\,u_0\,=\,0\,,\\\nonumber\\\nonumber
&& \lambda_{-1}\,=\,2\lambda\,w_1\,;\qquad\lambda_{-2}\,=\,-\lambda\,v_0\,;
\qquad\lambda_{-3}\,=\,\tfrac{5}{9}\lambda\,u_{-1}\,,
\eea
which then lead to nine independent Ward identities for nine currents,
\bea
&&\partial u_{-1}\,=\,2\lambda\,\bar\partial w_1\,,\qquad
\partial v_0\,=\,-3\lambda\,v_{-1}\,,\qquad
\partial v_{-1}\,=\,\lambda(-4v_{-2}\,+\,\tfrac{32}{5}\,v_0\,w_1)\,,
\label{eom}
\\\nonumber\\\nonumber
&&\partial v_{-2}\,=\,2\lambda\left(w_1\,v_{-1}\,+\,\tfrac{3}{5}v_0w_0\,-\,\bar\partial v_0\right)\,,\qquad 
\partial w_0\,=\,\lambda\left(\tfrac{10}{9}u_{-1}\,-\,4w_{-1}\,+\,8w_1^2\right)\,,
\\\nonumber\\\nonumber
&&
\partial w_1\,=\,-3\lambda\,w_0\,,\qquad \partial w_{-2}\,=\,\tfrac{4}{3}\lambda\left(u_{-1}w_1\,+\,4 w_1\,w_{-1}\,-\,4 v_0^2\,-\,6
w_{-3}\right),\\\nonumber\\\nonumber
&&
\partial w_{-1}\,=\,6\lambda w_0w_1\,-\,5\lambda w_{-2}\,,\qquad\partial\left(w_{-3}-\tfrac{3}{20}w_0^2 -\tfrac{1}{3}v_0^2 +\tfrac{2}{5}w_1w_{-1}\right)\,=\,\tfrac{5}{9}\lambda\bar\partial u_{-1}\,.
\eea
In the absence of the deformation $\lambda$ (which is required to be holomorphic in $z$), all nine currents are anti-holomorphic. The deformation 
introduces a holomorphic dependence in the currents and the Chern-Simons equations of motion can be viewed as anomalous Ward identities, induced by the deforming relevant operator in the UV CFT. The precise form of the above equations is determined by the OPEs between currents and the deforming operators.
 
Our task now is to demonstrate that the parameters above can indeed be mapped uniquely to a set of currents generating a $\cal W$-algebra associated to the 
$(2,1,1)$ non-principal embedding. 
This identification of the currents will proceed in steps:
\begin{itemize}
\item{ First, we outline the derivation of the ${\cal W}_4^{(2,1,1)}$ algebra from an $\mf{sl}(4)$ connection which is explicitly in the highest weight representation associated to the $(2,1,1)$ embedding (note this is different from what we actually have for the UV portion of \eqref{uvsl4}). This also makes explicit the form of such a connection in terms of the ${\cal W}$-algebra currents.}
\item{ We will show that an ${\rm SL}(4,\mathbb R)$ gauge transformation relates the UV portion of the gauge connection \eqref{uvsl4} to the highest weight decomposition  
in the $(2,1,1)$ embedding. This allows immediate identification of the mapping between the parameters $u_i, v_i, w_i$ and the currents of the ${\cal W}_4^{(2,1,1)}$ algebra.}
\item{Finally, we will show that the Chern-Simons equations of motion \eqref{eom} precisely reproduce the Ward identities for the ${\cal W}_4^{(2,1,1)}$ currents, with a unique choice for the UV stress tensor and the relevant operator perturbing the UV fixed point.}
\end{itemize}

\subsection{Asymptotic algebra for the $(2,1,1)$ embedding}

We begin by deriving the ${\cal W}$-algebra by considering an $\mf{sl}(4,\mathbb R)$ connection representing AdS$_3$ in the ``standard'' $(2,1,1)$ embedding:
\bea
&&A_{\rm AdS}\,=\,\hat L_1\,e^\rho\,dz + \hat L_0\, d\rho\,,\qquad\qquad
\bar A_{\rm AdS}\,=\,- \hat L_{-1}\,e^\rho\,dz - \hat L_0\, d\rho\,,
\label{connection3}\\\nonumber\\\nonumber
&&\hat L_1\,\equiv\,\tfrac{1}{6}W_3^{(4)}\,,\quad
\hat L_{-1}\,\equiv\,\tfrac{1}{6}W_{-3}^{(4)}\,,\qquad
\hat L_0\,=\,{\rm diag}\,\left(\tfrac{1}{2},0,0,-\tfrac{1}{2}\right)
\eea
Note that this gauge field is different from \eqref{uv2} and which describes the asymptotics of the black hole.  
 The generator of $\rho$-translations in that case is $L_0$ which differs from $\hat L_0$ by a certain $\U(1)$ generator. Nevertheless, as we show below in Sections \eqref{matching} and \eqref{wardid} using a Ward identity analysis, the asymptotic symmetry algebras for the two gauge connections are essentially the same.
For the sake of completeness, we outline the procedure determining the asymptotic symmetry algebra for the AdS background represented by eq.\eqref{connection3}, following the general idea in \cite{campoleoni}  (see also \cite{Afshar:2012hc}). 
In this approach, an asymptotically AdS connection must satisfy:\footnote{The following formulae are true for the embedding associated to $\hat L_1=W^{(N)}_{N-1}$. For a more general embedding, $\hat L_1=W^{(s)}_{s-1}$
it is necessary to take $\hat L_{-1}= U W^{(s)}_{-s+1}U^{-1}$ where
$U$ is an element of the Cartan subgroup, such that $\{\hat L_{\pm1}\}$ are generators of $\mf{sl}(2)$. Then the equation for the barred connection must be generalized slightly to include a gauge transformation $(U\bar AU^{-1}-\bar A_{\rm AdS})\big|_{\rho\to \infty}\,=\, {\cal O}(1)$.}
\be
(A-A_{\rm AdS})\big|_{\rho\to \infty}\,=\, {\cal O}(1)\,,\qquad\qquad
(\bar A-\bar A_{\rm AdS})\big|_{\rho\to \infty}\,=\, {\cal O}(1)\,.
\ee
The most general ${\rm SL}(4,\mathbb R)$ Chern-Simons connection, in the $(2,1,1)$ embedding, is (at a constant $\bar z$ slice) 
\bea
&&A_z\,=\sum_{n=-1}^1\,\hat{\cal L}_n(z)\,\hat L_n\,e^{n\rho}\,+
\,{\EuScript J}(z)\, J\,+
\sum_{A=-1}^1\,{\cal J}_A(z)\,J^A\,+\sum_{m=\pm \tfrac{1}{2}, a,b=\pm}{\cal G}_m^{ab}(z)\,G_m^{ab}\, e^{m\rho}\nonumber\\
&&A_{\bar z}\,=0\,, \label{general} 
\eea
and similarly for the barred connection. The 15 generators of $\mf{sl}(4)$ are split into a triplet $\{L_n\}$, four doublets $\{G_m^{ab}\}$ and four singlets
$\{J, J^A\}$ of $\mf{sl}(2)$:
\be
\underline {\bf 15} \to \underline{\bf 3}\,+\,4\cdot\underline{\bf 2}\,+\,4\cdot \underline{\bf 1}\,.
\ee
The explicit form for these generators is summarized in eq.~\eqref{gen} (Appendix \eqref{appa}). The four $\mf{sl}(2)$ singlets $\{J^A, J\}$ generate an $\text{SL}(2)'\times {\rm U}(1)$ symmetry.
We now impose the asymptotically AdS requirement by setting to zero the coefficients of all generators with positive weight under $\hat L_0$, and by fixing the coefficient of $\hat L_1$ to unity\footnote{It is not {\em a priori} clear how to directly apply the same procedure  to a connection representing the flow in eqs.~\eqref{uv1} and \eqref{uvsl4}, because dilatations ($\rho$-translations) are generated by $\tfrac{1}{3}L_0$, instead of $\hat L_0$ and hence the $\rho$-dependence is different from eq.~\eqref{asympads}.},
\be
\hat {\cal L}_1=1\,,\qquad\qquad{\cal G}_{1/2}^{ab}\,=\,0\,.
\ee
The constraint $\hat {\cal L}_1=1$ is first class and generates gauge transformations which can be used to set $\hat {\cal L}_0=0$. 
So we may take the gauge-fixed, asymptotically AdS connection to be, 
\be
A_z= \hat L_1\,e^\rho\,+
\,\EuScript J(z)J+
\sum_{A=-1}^1{\cal J}_A(z)\,J^A\,+\sum_{a,b=\pm} {\cal G}_{ab}(z)\,G_{-1/2}^{ab}\, e^{-\rho/2}+ \hat L_{-1}\,\hat{\cal L}(z)\,e^{-\rho}\,,
\label{asympads}
\ee
leaving nine undetermined functions which will eventually be identified with symmetry currents.   
In explicit matrix form, the asymptotically AdS, Chern-Simons gauge field is determined by the connection $ $
\bea
a_z=e^{\rho\hat L_0}A_z e^{-\rho\hat L_0}\,,\qquad
a_z\,=\,\left(\begin{array}{cccc}
\,\tfrac{1}{2}\EuScript J \,&\, {\cal G}_{++}\, &\, {\cal G}_{-+}\, &\, {\cal L}\\
&&&\\
0\,&\,-\tfrac{1}{2}\EuScript J+\tfrac{1}{2}{\cal J}_0 & {\cal J}_- \,&\, -{\cal G}_{--}\\
&&&\\
0 \,&\, - {\cal J}_+ \,&\,-\tfrac{1}{2}\EuScript J-\tfrac{1}{2}{\cal J}_0 \,&\,  {\cal G}_{+-}\\
&&&\\
1 \,&\, 0 \,&\, 0 \,&\,\tfrac{1}{2}\EuScript J
\end{array}
\right)
\label{asympz}
\eea
The asymptotic symmetries of this background are those ${\rm SL}(4,\mathbb R)$ gauge transformations which preserve the form of the gauge-fixed connection i.e. the gauge fixed connection, whilst not invariant, retains its specified form. The most general infinitesimal ${\rm SL}(4,\mathbb R)$ gauge transformations, holomorphic in $z$, are given by 
\bea
&&A_z\to A_z + \partial_z\Lambda\,+\,[A_z,\Lambda]\,,\\\nonumber
&& \Lambda\,=\,\sum_{n=-1}^1\,\varepsilon_n(z)\,\hat L_n\,+
\,\gamma(z)\, J\,+
\sum_{A=-1}^1\,\eta_A(z)\,J^A\,+\sum_{m=\pm 1/2, a,b=\pm} \chi_m^{ab}(z)\,G_m^{ab}\,.
\eea
These transformations preserve the form of the AdS connection only if the gauge parameters satisfy certain relations,
\bea
&&\varepsilon_0\,=\,-\varepsilon_1'\,,\\\nonumber\\\nonumber
&&\varepsilon_{-1}\,=\,\tfrac{1}{2}\varepsilon_1'' - \hat{\cal L}\,\varepsilon_1+\tfrac{1}{2}{\cal G}_{--}\,\chi_+^{++}\,+\tfrac{1}{2}{\cal G}_{-+}\,\chi_+^{+-}\,
-\tfrac{1}{2}{\cal G}_{+-}\,\chi_+^{-+}\,-\tfrac{1}{2}{\cal G}_{++}\,\chi_+^{--}\,,\\\nonumber\\\nonumber
&&\chi_-^{-+}\,=\,-\chi_+^{-+ \,'}+\varepsilon_1\,{\cal G}_{-+}+ {\cal J}_-\,\chi_+^{++}-\tfrac{1}{2}{\cal J}_0\,\chi_+^{-+}-\EuScript J\,\chi_+^{-+}\,,\\\nonumber\\\nonumber
&&\chi_-^{+-}\,=\,-\chi_+^{+- \,'}+\varepsilon_1\,{\cal G}_{+-}- {\cal J}_+\,\chi_+^{--}+\tfrac{1}{2}{\cal J}_0\,\chi_+^{+-}+\EuScript J\,\chi_+^{+-}\,,\\\nonumber\\\nonumber
&&\chi_-^{--}\,=\,-\chi_+^{-- \,'}+\varepsilon_1\,{\cal G}_{--}+ {\cal J}_-\,\chi_+^{+-}-\tfrac{1}{2}{\cal J}_0\,\chi_+^{--}+\EuScript J\,\chi_+^{--}\,,\\\nonumber\\\nonumber
&&\chi_-^{++}\,=\,-\chi_+^{++ \,'}+\varepsilon_1\,{\cal G}_{++}- {\cal J}_+\,\chi_+^{-+}+\tfrac{1}{2}{\cal J}_0\,\chi_+^{++}-\EuScript J\,\chi_+^{++}\,.
\eea
Hence there are nine independent gauge transformations generated by 
gauge parameters $\varepsilon_1,\gamma, \eta_a, \chi_+^{ab}$ which preserve the form of the asymptotically AdS connection. It is straightforward to deduce the variations of all currents under the  independent gauge transformations. The variations of the currents are listed in eqs.~\eqref{var1}-\eqref{var4}.
The gauge transformations of the currents completely determine the asymptotic symmetry algebra since the Poisson brackets of the charges can be read off directly from the above transformations. Specifically, the symmetry variation of any phase space functional is given by \cite{campoleoni}
\be
\delta_\Lambda F\,=\,\{Q(\Lambda), F\}\,,\qquad Q(\Lambda)\,\equiv\,-\frac{k}{2\pi}\int d\phi\,{\rm Tr}\,(a_\phi\,\Lambda^+)\,,
\ee
where $\Lambda^+$ picks out the components of $\Lambda$ with non-negative weights with respect to $\hat L_0$, and it is understood that the charges are computed on a constant time slice. Explicitly, we obtain
\bea
&&Q(\Lambda)\,=-\tfrac{k}{2\pi}\left(\,{\cal G}_{++}\,\chi^{--}\,+\,
{\cal G}_{+-}\,\chi^{-+}\,-\,{\cal G}_{-+}\,\chi^{+-}\,-\,
{\cal G}_{--}\,\chi^{++}\,+\,\hat{\cal L}\,\varepsilon_1\,+\EuScript J\,\gamma\,\nonumber\right.\\
&&\left.\qquad\qquad+\tfrac{1}{2}{\cal J}_0\,\eta_0\,-\,{\cal J}_+\,\eta_-\,-\, {\cal J}_-\eta_+\right)\,.
\eea
The resulting Poisson brackets for the ${\cal W}$-algebra generated by the currents are listed in \eqref{PB}. It is important to note that to recover the  correct dependence of the central charge on the Chern-Simons level, we must correctly normalize the currents, which we can do by performing the rescalings
\bea
\hat{\cal L} \to -\tfrac{2\pi}{k}\hat{\cal L}\,,\quad {\cal G}_{ab}\to
-\tfrac{2\pi}{k}{\cal G}_{ab}\,,\quad{\cal J}_\pm\to-\tfrac{2\pi}{k}{\cal J}_\pm\,,\quad {\cal J}_0\to-\tfrac{4\pi}{k}{\cal J}_0\,,\quad \EuScript J \to -\tfrac{2\pi}{k}\EuScript J\ .\nonumber
\eea
It also becomes clear from the Poisson brackets that in order for the spin-$\tfrac{3}{2}$ currents to have a regular tensor transformation law, the naive stress tensor ${\cal L}$ needs to be modified, 
\be
\hat{\cal L}\,\to\, \boxed{T\,=\,\hat{\cal L}-\tfrac{2\pi}{k}\left(\tfrac{1}{2}\EuScript J^2+
{\cal J}_0^2 -{\cal J}_+\,{\cal J}_-\right)}\ .
\label{improve1}
\ee
\\
We recognize this modification as the Sugawara form for the stress 
tensor of the $\mf{sl}(2)'\oplus\mf{u}(1)$ affine algebra.
Since we are working in a semiclassical limit $(k\to \infty)$, terms non-linear in the currents appear with a power of $k^{-1}$, and in addition we can ignore normal ordering effects from such terms.
The Poisson brackets imply the following non-trivial OPEs for the currents of the ${\cal W}_{4}^{(2,1,1)}$-algebra:
\bea
&&T(z)\,T(0)\,\sim\,\frac{3k}{z^4}\,+\,\frac{2}{z^2}\,T\,+\,\frac{1}{z}\partial T\,,\qquad T(z)\,{\cal G}_{ab}(0)\,\sim\,\frac{3}{2z^2}{\cal G}_{ab}\,+\,\frac{1}{z}\partial {\cal G}_{ab}\,,\nonumber\\\label{opes}\\\nonumber
&&
T(z)\,{\cal J}_{A}(0)\,\sim\,\frac{1}{z^2}{\cal J}_{A}\,+\,\frac{1}{z}\partial {\cal J}_{A}\,,\qquad\quad
T(z)\,{\EuScript J}(0)\,\sim\,\frac{1}{z^2}\EuScript J\,+\,\frac{1}{z}\partial \EuScript J\,,
\eea
\bea
&& {\cal J}_0(z)\,{\cal J}_0(0)\,\sim\,-\frac{k}{2z^2}\,,
\qquad {\cal J}_+(z)\,{\cal J}_-(0)\,\sim\,\frac{k}{z^2} -\frac{2}{z}{\cal J}_0\,,
\qquad 
{\cal J}_0(z)\,{\cal J}_{\pm}(0)\,\sim\,\pm\frac{1}{z}\,{\cal J}_{\pm}(0)\,,\nonumber\\\nonumber\\\nonumber
&&\EuScript J(z)\,\EuScript J(0)\,\sim\,-\frac{k}{z^2}\,,
\qquad\qquad {\cal G}_{\pm \mp}(z)\,{\cal G}_{\pm\pm}(0)\,\sim\,
\frac{2}{z^2}{\cal J}_\pm\,+\,\frac{1}{z}\partial{\cal J}_\pm\mp\,
\frac{2}{k\,z} \EuScript J\,{\cal J}_\pm\,,\\\nonumber\\\nonumber
&&\EuScript J(z)\,{\cal G}_{a\pm}(0)\,\sim\,\mp\frac{{\cal G}_{a\pm}}{z}\,,\qquad {\cal J}_0(z)\,{\cal G}_{\pm a}(0)\,\sim\,\pm\frac{1}{2z}{\cal G}_{\pm a}\,,\qquad
{\cal J}_\mp(z)\,{\cal G}_{\pm a}(0)\,\sim\,\pm\frac{1}{2 z}{\cal G}_{\mp a}\,,\\\nonumber\\\nonumber
&&{\cal G}_{-\mp}(z)\,{\cal G}_{+\pm}(0)\,\sim\,\frac{2k}{z^3}
\,+\,\frac{2}{z^2}\left({\cal J}_0\mp\EuScript J\right)\,+\,\frac{2}{z}\left(\partial{\cal J}_0\mp\partial\EuScript J+\,\tfrac{1}{2}T\right)\,\\\nonumber
&&\hspace{2.3in}\,+\,
{k^{-1}}\,\frac{1}{z}\left(2{\cal J}_+{\cal J}_- - 2{\cal J}_0^2\pm \EuScript J{\cal J}_0-\tfrac{3}{2}\EuScript J^2\right)\,.
\eea
The currents of the ${\cal W}$-algebra comprise of the stress tensor $T$, four $\mf{sl}(2)'\oplus\mf{u}(1)$ currents and four spin-$\tfrac{3}{2}$ currents ${\cal G}_{ab}$. The central charge is $\hat c\,=\,6k$, which is precisely what we expect from a two dimensional embedding of $\mf{sl}(2)$. The levels of the $\U(1)$ currents ${\cal J}_0$ and $\EuScript J$ are negative, and the OPEs contain non-linear terms that are suppressed by $k^{-1}$ in the large-$k$ regime, which is also the regime in which the classical Chern-Simons description is valid. The algebra agrees with the semiclassical limit of the result quoted in \cite{Afshar:2012hc}.

\subsection{Matching black hole parameters to ${\cal W}$-algebra currents}
\label{matching}
Having determined the explicit form of the ${\cal W}$-algebra corresponding to the standard $(2,1,1)$ embedding, we turn our attention to the $\mf{sl}(4)$ gauge connection \eqref{uvsl4} in the presence of the spin-4 chemical potential.  
Remarkably, we note that the anti-holomorphic component, $a_{\bar z}$ of eq.~\eqref{uvsl4} can be put in the form of eq.~\eqref{asympz}, using an ${\rm SL}(4,\mathbb R)$ gauge transformation:
\be
a_{\bar z}\,\to\, \Omega^{-1}\,\left(a_{\bar z}+\partial_{\bar z}\right)\,\Omega\,,
\qquad\qquad\Omega\,=\,\frac{1}{10}\,\left(\begin{array}{cccc}
\,1 \,&\, 0\, &\,-4\sqrt{3}\, w_1\, &\, -3\,w_0\\
\,0\,&\,1\, &\,0 \,&\,4\sqrt{3}\,w_1 \\
\,0 \,&\, 0 \,&\,1 \,&\, 0 \\
\,0 \,&\, 0 \,&\, 0 \,&\,1
\end{array}
\right),
\label{gauge}
\ee
which results in a dictionary between the parameters of the black hole gauge connection and the ${\cal W}$-algebra currents, that we quote in eq.~\eqref{dictionary}.  This is a non-trivial fact because it indicates that, given the IR piece $a_z$ of \eqref{uvsl4} in highest weight gauge in the principal embedding, the Chern-Simons flatness conditions force the UV piece $a_{\bar z}$ to be gauge equivalent to a highest weight decomposition in the $(2,1,1)$ embedding.

With these identifications in place the ${\rm SL}(4, \mathbb R)$ Chern-Simons equations \eqref{eom} can be rewritten in terms of anti-holomorphic ${\cal W}$-algebra currents.
Noting that precisely the same identifications hold for the Chern-Simons connection in the barred sector, we list the equations obeyed by the holomorphic currents which acquire anti-holomorphic pieces due to the UV deformation,
\bea
&&\bar\partial\tilde T\,=\,\tfrac{1}{3}\,\lambda\,\partial{\cal O}\,,\qquad\bar\partial{\cal J}_+\,=\,4\lambda\,{\cal J}_0\,,\qquad\bar\partial\EuScript J\,=\,\sqrt 3\lambda\,\left({\cal G}_{++}\,-\,{\cal G}_{+-}\right)\\\nonumber\\\nonumber
&&\bar\partial{\cal O}\,=\,-\tfrac{50}{3}\lambda\,\partial{\cal J}_+\,,\qquad\bar\partial{\cal J}_0\,=\,-\tfrac{1}{2}\lambda\,\left({\cal O}-6{\cal J}_-+ 8{\cal J}_+^2\right)\\\nonumber\\\nonumber
&&\bar\partial{\cal J}_-\,=\,-\lambda\,
\left\{\sqrt{3}({\cal G}_{--}+{\cal G}_{-+})+
\tfrac{16}{3}\left(\partial{\cal J}_+ \,+\,
2{\cal J}_0{\cal J}_+ \right)\right\}\\\nonumber
\\\nonumber
&&\sqrt3\,\bar\partial\left({\cal G}_{++}-{\cal G}_{+-}\right)\,=\,-\lambda\,\left\{2\sqrt3\,\left({\cal G}_{--}\,-\,{\cal G}_{-+}\right)\,+\,12\,\EuScript J\,{\cal J}_+\right\}\\\nonumber\\\nonumber
&&\sqrt3\,\bar\partial\left({\cal G}_{-+}-{\cal G}_{--}\right)\,=\,-2\lambda\,\left\{\tfrac{8}{\sqrt 3}\,\left({\cal G}_{++}\,-\,{\cal G}_{+-}\right)\,{\cal J}_+\,+\,3\left(2{\cal J}_0\EuScript J\,+\,\partial\EuScript J\right)\right\}
\eea
where the improved stress tensor $\widetilde T$ and the operator ${\cal O}$ are defined as
\\
\bea
\boxed{{\widetilde T}\,=\,T\,+\,\tfrac{1}{3}\partial{\cal J}_0\,}\ ,\qquad
\boxed{{\cal O}\,=\,\sqrt 3\left({\cal G}_{++}\,+\,{\cal G}_{+-}\right)\,+\,2{\cal J}_-\,+\,\tfrac{8}{3}{\cal J}_+^2\,}\ .
\label{improve2}
\eea
\\
This ``improved" stress tensor is forced upon us as the natural combination that appears in the dictionary \eqref{dictionary} between the parameters of the black hole connection and the ${\cal W}$-algebra currents. 
All currents are holomorphic in the absence of the deformation parameter $\lambda$.
To summarize, we have re-expressed the Chern-Simons equations of motion in terms of 
currents of the ${\cal W}_4^{(2,1,1)}$-algebra. It remains to show that these equations are 
precisely Ward identities in the theory obtained by perturbing the UV fixed point by the relevant operator ${\cal O}$.

 \subsection{Matching to ${\cal W}_4^{(2,1,1)}$ Ward identities}
 \label{wardid}
We have seen that the asymptotic form of the $\text{SL}(4,\mathbb R)$ Chern-Simons connection \eqref{connection1} describes the introduction of a chemical potential for spin-4 charge in the ${\cal W}_4$ CFT. The corresponding deformation is irrelevant from the perspective of this (IR) conformal fixed point. The connection, however, describes a flow to a new (UV) AdS geometry. From the background metric, we infer that the RG flow results from perturbing the UV fixed point by an operator of dimension $4/3$ as measured by the stress tensor whose zero mode is $\hat L_0 +\tfrac{1}{3} J_0$. Therefore, by inspecting the ${\cal W}_4^{(2,1,1)}$ algebra above, we conclude that the UV stress tensor appropriate for describing the RG flow background is 
\\
\be
\boxed{{\widetilde T}\,=\,T\,+\,\tfrac{1}{3}{\cal J}'_0}\ .
\ee
\\
There are precisely three operators with dimension $4/3$ (as measured by $\widetilde T$) at the UV fixed point, hence we write the perturbing operator as
\be
{\cal O}\,=\,g_+\,{\cal G}_{++}\,+\,g_-\,{\cal G}_{+-}\,+
g_1\,{\cal J}_-\,+\,g_2\,{\cal J}_+^2\,.
\ee
Once the UV field theory is deformed by this operator,
\be
I_{\rm UV}\,\to\,I_{\rm UV} +\lambda\int d^2 z\,\left({\cal O}(z)\, +\,\overline{\cal O}(\bar z)\,\right),
\ee
all CFT currents obey ``Ward identities'' that are completely determined by their Poisson brackets/OPEs with the deforming operator ${\cal O}$. Currents which were holomorphic at the fixed point acquire anti-holomorphic pieces (and vice-versa) along the flow, governed by their respective OPE's with ${\cal O}$.  Making use of the identity $\partial_{\bar z}\left(\tfrac{1}{z}\right)\,=\,2\pi\delta^2(z,\bar z)$, to linear order in $\lambda$, using the OPEs for the ${\cal W}$-algebra, we find the Ward identities of the perturbed UV conformal fixed point:
\bea
&&\langle\bar\partial {\tilde T}\rangle\,=\,\frac{\lambda}{3}\,\langle\partial 
{\cal O}\rangle\,,\qquad \langle\bar\partial {\cal J}_+\rangle\,=\,2\lambda\,g_1\,\langle{\cal J}_0\rangle\,,\qquad \langle\bar\partial \EuScript J\rangle\,=\,\lambda\langle\left(g_+\,{\cal G}_{++}
\,-\,g_-\,{\cal G}_{+-}\right)\rangle\,,\nonumber\\\nonumber\\\nonumber
&&\langle\bar \partial {\cal O}\rangle\,=\,-2\lambda\,
\langle\left(g_1g_2\, +\, g_+ g_-\right)\,\partial {\cal J}_+\rangle\,,\qquad\,
\langle\bar \partial {\cal J}_0\rangle\,=\,-\tfrac{1}{2}\lambda\,\langle\left({\cal O}\,-\,3 g_1\,{\cal J}_-\,+\,3g_2\,{\cal J}_+^2\right)\rangle\,,\\\nonumber\\
&&\langle\bar\partial {\cal J}_-\rangle\,=\,-\lambda\,\langle\left(
g_+\,{\cal G}_{-+}\,+\,g_-\,{\cal G}_{--}
\,+\,2g_2\,\partial {\cal J}_+ 
\,+\,4g_2\,{\cal J}_0{\cal J}_+\right)\rangle\,,\\\nonumber\\\nonumber
&&\langle\bar\partial(g_+\,{\cal G}_{++}\,-\,g_-\,{\cal G}_{+-})\rangle
\,=\,-\lambda\langle\left\{4g_+g_-\,\EuScript J\,{\cal J}_+
\,-\,g_1\,(g_+{\cal G}_{-+}\,-\,g_-{\cal G}_{--})
\right\}\rangle\,,\\\nonumber\\\nonumber
&&\langle\bar\partial(g_+{\cal G}_{-+}\,-\,g_-{\cal G}_{--})\rangle\,=\,
-2\lambda\langle\left\{g_+g_-(\partial \EuScript J+ 2\EuScript J{\cal J}_0)
\,+\,g_2{\cal J}_+(g_+{\cal G}_{++}-g_-{\cal G}_{+-})
\right\}\rangle\ .
\eea
Comparing with the Chern-Simons equations of motion we immediately notice beautiful agreement once the deformation parameters are fixed to
\be
g_+\,=\,g_-\,=\,\sqrt 3\,,\qquad g_1\,=\,2\,,\qquad
g_2\,=\,\frac{8}{3}\,.
\ee
Using the OPEs \eqref{opes} the central charge as measured by the improved stress tensor $\tilde T$ is given by
\be
c_{\rm UV}\,=\,\frac{20\,k}{3}\,.
\ee
On the other hand, the central charge of the ${\cal W}_4$ CFT in the IR is given by
\be
c_{\rm IR}\,=\,12\,k\,{\rm Tr}(L_0^2)\,=\,60k\,.
\ee
Therefore, $c_{\rm UV}/c_{\rm IR} = 1/9$, as expected from the 
ratio  of the respective AdS-radii \eqref{ratioc}. Note that this does not violate the c-theorem because Lorentz invariance is not preserved by the flow.
\subsection{Summary}
In this section we have demonstrated by explicit computation, that the Chern-Simons connection representing a spin-4 black hole in ${\rm SL}(4,\mathbb R)$ Chern-Simons theory, also describes a flow between two CFT's characterized by different ${\cal W}$-algebras. In particular, we have shown that the UV conformal fixed point has a ${\cal W}$-algebra  associated to the $(2,1,1)$ non-principal embedding of $\mf{sl}(2)$ in $\mf{sl}(4)$. Crucially, conformal transformations of this UV CFT are generated by a twisted stress tensor $\tilde T$. Our analysis strongly suggests that an analogous picture will continue to be valid for general higher spin black holes in ${\rm SL}(N,\mathbb R)$ and that these will correspond  to flows between CFT's associated to the 
principal and some non-principal embedding of $\mf{sl}(2)$ in $\mf{sl}(N)$.

\section{Thermodynamics}

We have presented evidence that some of the key features of higher-spin black hole backgrounds in AdS$_3$, originally found in \cite{gutkraus} for ${\rm SL}(3,\mathbb R)$ Chern-Simons theory, can be naturally generalized to ${\rm SL}(N,\mathbb R)$ theories with $N> 3$. In particular the black holes with higher spin charge continue to be embedded within flows between CFT's characterized by ${\cal W}$-algebras corresponding to different embeddings of $\mf{sl}(2)$ in $\mf{sl}(N)$. It is now interesting to ask whether the thermodynamics of these black holes and in particular the interplay between multiple branches of solutions \cite{David:2012iu} also generalizes in a simple way.

\paragraph{\underline{Thermodynamic action}:} Our approach towards the thermodynamics will first follow the general idea outlined in \cite{bct}, which was explicitly applied to spin-three black holes in \cite{David:2012iu}. As pointed out in \cite{deboer} this so-called ``holomorphic" formulation of thermodynamics differs from the ``canonical" formulation followed in \cite{deboer} which coincides with the approaches of \cite{perez} and \cite{Campoleoni:2012hp}\footnote{The result of \cite{Campoleoni:2012hp} is noteworthy in that the entropy (which can now be recognized as the one in the ``canonical'' formulation) of the higher spin black hole for weak spin-3 field was computed by means of Wald's formula \cite{Wald:1993nt} as a correction to the area law.}.

In order to be identified with black holes, the $\mf {sl}(4)$ connections in eq.~\eqref{sl4}, must also satisfy a smoothness condition in Euclidean signature. In particular, the holonomy of the gauge field around the Euclidean thermal circle must be trivial. This is analogous to the requirement of regularity of the Euclidean black hole metric  in ordinary gravity \cite{gutkraus, ammon, conical}. The smoothness condition in turn implies that the holonomy of the gauge field should lie in the center of the 
gauge group, which then constrains the eigenvalues of the connection $a_t$. To describe a higher spin black hole which is connected smoothly to the usual BTZ black hole at $\mu=0$, we require  
\be
\exp\left(\oint_\beta a_t\right)\,=\,\exp\left(2\pi i L_0\right)\,,\qquad
a_t\,\equiv\,i(a_z+a_{\bar z})\ .
\ee
This condition fixes the eigenvalues of the matrix $\beta a_t$, to match those of the dilatation generator $L_0\,\equiv\,{\rm diag}(\,\tfrac{3}{2},\, \tfrac{1}{2},\,-\tfrac{1}{2},\,-\tfrac{3}{2} )$, so that
\be
{\rm Tr}(a_t^2)\,=\,-20\pi^2T^2\,,\qquad{\rm Tr}(a_t^3)\,=\,0\,,\qquad
{\rm det}(a_t)\,=\,9\pi^4 T^4\,.
\ee
Requiring that the holonomies be fixed, automatically guarantees that the first law of thermodynamics is obeyed by the solutions \cite{deboer,conical2}. 
In the approach of \cite{bct} the on-shell Euclidean action of the Chern-Simons theory reads
\be
I_{\rm on-shell}\,=\,\frac{i k}{4\pi}\,\int dt\,d\phi\,\left[
\,{\rm Tr} (a_t \,a_\phi)\quad-\quad{\rm Tr} (\bar a_t \,\bar a_\phi)\, \right]\,.
\ee
This is purely a boundary contribution (in the so-called angular quantization picture) since the bulk Chern-Simons action simply vanishes on-shell.
On the other hand, a free variation of the Chern-Simons bulk and boundary terms yields
\be
\delta I\,=\,-\frac{ik}{4\pi}\int_{{\mathbb T}^2}dt\,d\phi\,
\left[\,{\rm Tr}
\left(a_\phi\,\delta a_t-a_t\,\delta a_\phi\right)\quad - \quad
(a\to \bar a)\,\right]\ ,
\label{deltacs}
\ee
where the variation includes the effect of changing the modular parameter of the boundary torus, or the inverse temperature $\beta$. Assuming a non-rotating configuration, explicit evaluation of this using the connections \eqref{sl4} (or those in \cite{David:2012iu} for ${\rm SL}(3, \mathbb R)$) shows that $I$ does not transform as a thermodynamic grand potential. 
\be
\delta I\,=\,- 8\pi Q_4\,\beta\,d\mu - 4\pi dQ_4\,(\beta\mu) - \,d\beta\,(I_{\rm on-shell}\,\beta^{-1})\,.
\ee
This can, however, be fixed by a Legendre transform of $I$, 
\be
I \to  I \,+\, 4\pi{Q_4}\,(\beta\mu)\,.
\ee
It is now straightforward to deduce that the correct thermodynamical action 
which yields the expected variations with respect to the inverse temperature 
$\beta$ and the chemical potential $\mu$,
\be
\frac{\partial {I_{\rm th}}}{\partial \beta}\,=\,4\pi{\cal L}\,,\qquad\qquad
\frac{\partial { I_{\rm th}}}{\partial(\beta\mu)}\,=\,-\,4\pi\,Q_4\,,
\ee
is given by,
\be
{\Phi}\,\equiv\,\beta^{-1}\,I_{\rm th}\,=\, -4\pi {\cal L}\,+\,12\pi\mu\,Q_4\ .
\ee
$\Phi$ is the grand potential, and we have assumed that the barred and unbarred 
sectors contribute equally to the total energy and spin-four charge, for the non-rotating configuration. From this we deduce the thermodynamic entropy 
\be
S\,=\,\frac{1}{T}\left(8\pi {\cal L} -16\pi\,\mu\,{Q}_4\right)\,.
\ee
We may also readily verify, that the formulae for the entropy and the grand potential 
can be directly deduced from the Chern-Simons connections as
\bea
&&\Phi\,=\, -k\,{\rm Tr}\left(a_z^2\,+\,\tfrac{3}{2}\,a_z\,a_{\bar z}\right)\,,\label{xx1}\\\nonumber
\\\nonumber
&&S\,=\,2ik\,\beta\,{\rm Tr}(a_t\,a_z)\,=\,2k\,\beta\,{\rm Tr}\left(a_z^2\,+\,a_z\,a_{\bar z}\right)\,.
\eea
 This expression for entropy was pointed out in \cite{deboer, conical2}. In particular, in \cite{conical2}, it was shown that this entropy formula (for non-rotating solutions) can be obtained by evaluating the Chern-Simons action ``off-shell'' on appropriately regularized  singular connections, i.e. those with non-trivial holonomy around the Euclidean time circle. 

Note that our expressions for the entropy and the energy differ from those in the 
canonical formulation where the energy includes terms non-linear in ${\cal L}$. In the holomorphic formulation we adopt, the energy is naturally defined as the expectation value of the Hamiltonian of the undeformed theory.

\subsection{Holonomy conditions and phase structure}
In \cite{David:2012iu}, the potentially rich thermodynamic phase structure of the spin-3 black hole solutions in ${\rm SL}(3, {\mathbb R})$ Chern-Simons theory was shown. The main feature was the existence of multiple branches of black hole solutions of which only one is smoothly connected to the BTZ black hole at $\mu=0$. Perhaps the most remarkable feature of the thermodynamics of spin-3 black holes is that the BTZ branch ceases to exist beyond a critical value of the chemical potential (at fixed temperature).\footnote{One possible resolution is that beyond this point one must allow for complex roots of the holonomy conditions, since ${\rm  SL}(3,\mathbb C)$ is the true gauge group of the Euclidean theory. In that case the BTZ-branch continues to exist beyond this critical chemical potential, as a complex saddle point, until eventually at some higher $\mu$ there is a first order transition to the phase that dominates in the UV.}
At this critical point, the BTZ branch merges with a thermodynamically unstable branch. Finally, the physics at ultra-high temperature  (or fixed temperature, ultra-high chemical potential) is remarkably well described by the thermodynamics of a UV CFT with ${\cal W}_3^{(2)}$ symmetry.
 
It is now easy to verify whether this thermodynamic phase structure generalizes in a simple way to ${\rm SL}(N,\mathbb R)$ theories with $N=4,5$. Black hole solutions require trivial holonomy of the Chern-Simons connection \eqref{sl4} around the time circle. For $N=4$, we obtain three algebraic equations for the three charges
${\cal L}$, $Q_3$ and $Q_4$. The equations have multiple roots, with six real branches (out of a total of 27 roots). Remarkably, the phase diagram bears a strong similarity to the ${\rm SL}(3, {\mathbb R})$ theory.  Once again, the BTZ branch ceases to exist beyond a critical value of $\sqrt \mu\,  T$. 
 
\begin{figure}[h]
\begin{center}
\epsfig{file=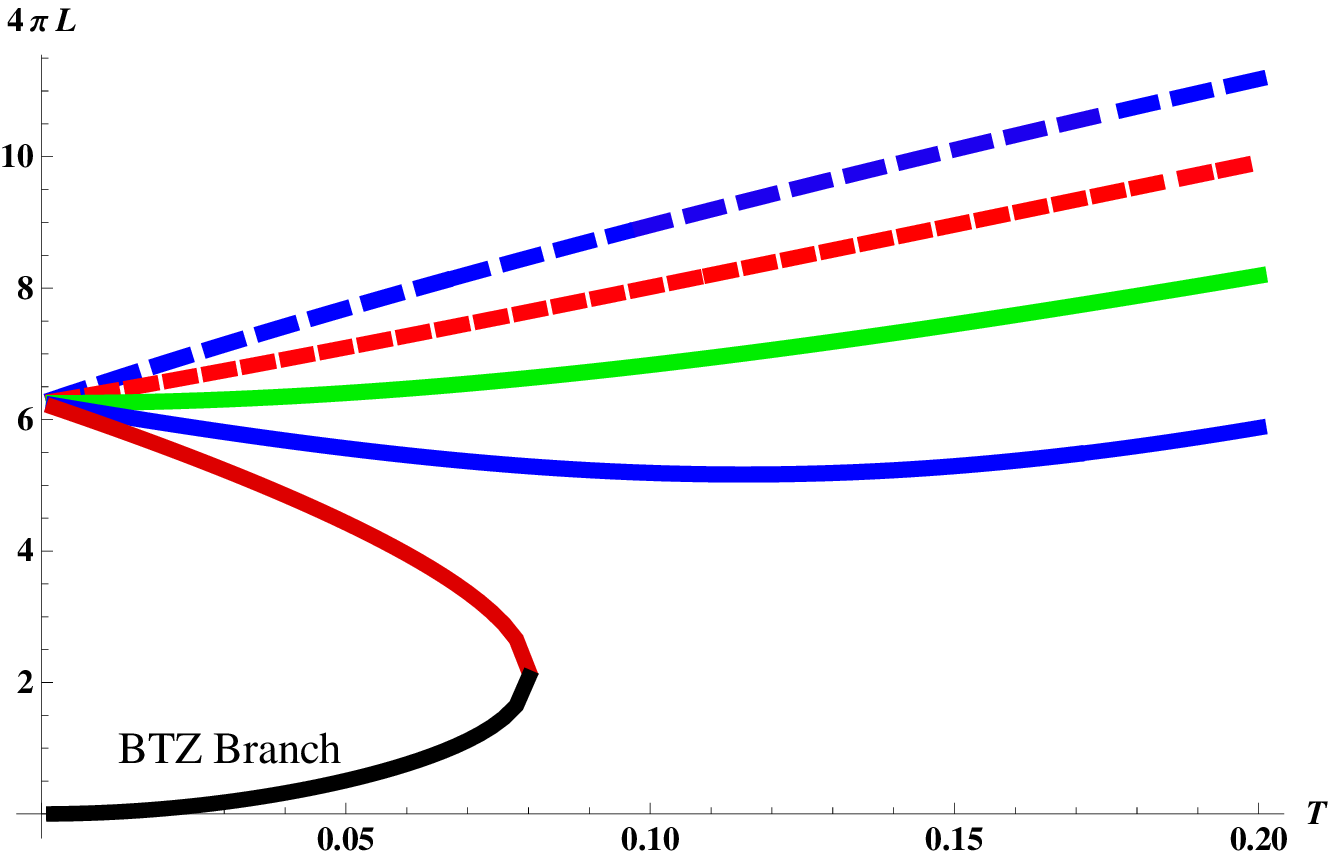, width =2.3in}\hspace{1in}
\epsfig{file=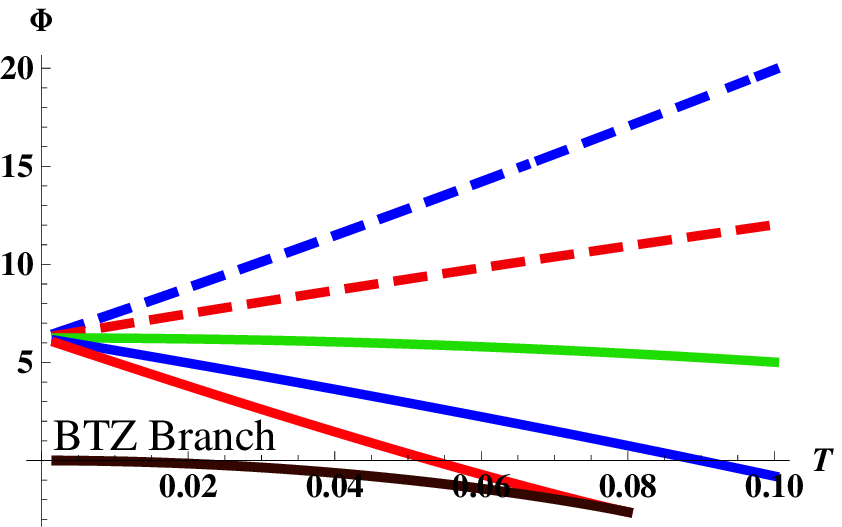, width =2.3in}
\end{center}
\caption{\small{ 
Energy and the grand potential for the six branches corresponding to 
spin-4 black holes in ${\rm SL}(4,{\mathbb R})$ Chern-Simons theory.
}}
\label{thermod4} 
\end{figure}
Performing the analogous excercise for ${\rm SL}(5, {\mathbb R})$ Chern-Simons theory, we find a much larger number of roots, ranging between 46 real roots for small $T$ and  $18$ real branches for larger temperatures (at fixed $\mu$).
\begin{figure}[h]
\begin{center}
\epsfig{file=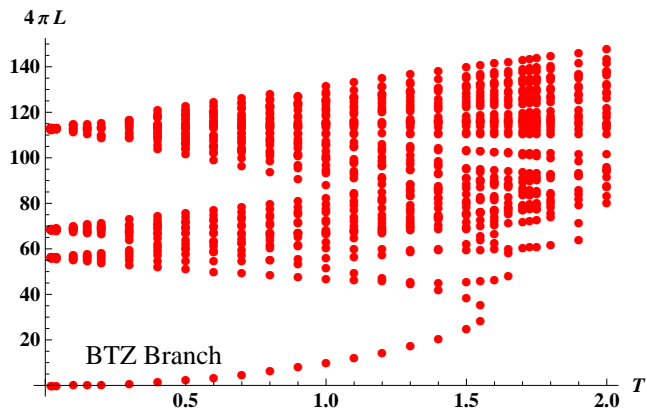, width =3.3in}\hspace{1in}
\end{center}
\caption{\small{ 
Energy versus temperature for real  branches corresponding to 
spin-5 black holes from ${\rm SL}(5,{\mathbb R})$ Chern-Simons theory.
}}
\label{thermod5} 
\end{figure}
The universal features of the solutions are self-evident. In particular, the disappearance of the BTZ-branch is a robust feature. The other striking feature is the rapid growth in the number of solutions with $N$. This is due to the fact that there are $N-1$ holonomy conditions on the $N-1$ thermodynamical charges and the order of the polynomial equations arising from the holonomy conditions effectively grows as $\sim N^2$ for large $N$. It would be a very interesting excercise to uncover the large-$N$ scaling of the number of real branches and whether the rapid growth of this number has implications for the presence or absence of phase transitions in the theory with an infinite tower of higher spin fields.

\subsection{Thermodynamics of the UV fixed point}
Finally we turn to a non-trivial consistency check of the picture we have presented thus far.
This is provided by the construction of a thermodynamical action for the UV fixed point  in terms of the appropriate variables for that theory.
This is once again analogous to, and a generalization of the situation with the spin-3 black hole, discussed in \cite{David:2012iu}. To obtain the UV thermodynamics, we rewrite the flat connections, after a constant gauge transformation (rescaling),
\bea
&& \hat a\,=\,\left(\tfrac{1}{6}\,W_{3}^{(4)}\,+\,w_{1}\,W_{1}^{(4)}
\,+\,w_{-3}\,W_{-3}^{(4)}\,+\,w_{-1}\,W_{-1}^{(4)}\,+
\,v_{-2}\,W_{-2}^{(3)}\,+ \right.
\\\nonumber\\\nonumber 
&&\left.
+ \, v_{0}\,W_{0}^{(3)} \,+\,
u_{-1}\, L_{-1}\right)
\,d\bar z\,+\,\lambda\left(L_1\,+\,2w_1\, L_{-1}\,- \,v_0\,W_{-2}^{(3)}\,+\,\tfrac{5}{9}u_{-1}\,W_{-3}^{(4)}\right)\,dz\,,
\eea
where we have swapped the $z$ and $\bar z$ coordinates. 
Now we can compute the complete variation of the action and deduce thermodynamic quantities
following the steps already outlined above. We calculate the variation of the Chern-Simons action \eqref{deltacs} after a suitable Legendre transform, using the equations of motion \eqref{eom} and the relation between Chern-Simons variables and {\cal W}-algebra currents \eqref{dictionary}, and we find,
\be
\delta \hat I\,=\,- 4\pi\, \langle \tilde T\rangle\,d\beta\,- 4\pi\,\langle{\cal O}\rangle\,d(\lambda\beta)\,.
\ee
$\langle\tilde T\rangle$ and $\langle\cal O\rangle$ are the expectation values of the UV stress tensor
and the deforming current (\eqref{improve1},\eqref{improve2}), obtained from the asymptotic symmetry algebra and the Ward identities. Therefore the energy as measured with respect to the UV fixed point coincides with the expectation value of the stress tensor obtained using different methods, and $\lambda$ can be viewed as a chemical potential for the conserved current ${\cal O}$ (scaling dimension $\tfrac{4}{3}$). Note that in Euclidean signature, the energy density is proportional to $- \langle\widetilde T\rangle$. The grand potential for this ensemble is 
\be
\hat\Phi\,=\,4\pi\,\langle{\widetilde T}\rangle\,+\,4\pi\lambda\,
\langle{\cal O}\rangle\,=\,
-k\,{\rm Tr}\left(\hat a_z^2 \,+\,\tfrac{3}{2}\,\hat a_z\,\hat a_{\bar z}\right)\,.\label{xx2}
\ee
Before we leave the discussion of the thermodynamics, the outstanding issue that we do not solve here is to provide a definition of the free energy which interpolates smoothly between the IR and UV definitions we have provided above in \eqref{xx1} and \eqref{xx2}, respectively.

\section{Discussions and Conclusions}\label{dis}

The goal of this work was to analyze in detail the asymptotic symmetries of higher spin black hole solutions in ${\rm SL}(N, \mathbb R)$ Chern-Simons theory. Although we restricted attention to the case with highest spin chemical potential for simplicity, our analysis confirms the general expectation that these solutions should correspond to flows linking a ${\cal W}_N$ CFT to one with a non-principal ${\cal W}$-algebra. To be more specific, we conjecture that the UV limit 
associated to the element $W^{(s)}_{s-1}$ will involve the ${\cal W}$-algebra for the $\mf{sl}(2)$ embedding with $\hat L_1=W^{(s)}_{s-1}$. The only 
subtlety in this identification is that the stress tensor of the UV fixed point must be appropriately twisted by $\U(1)$ currents which always arise in non-principal embeddings of $\mf {sl}(2)$ in $\mf{sl}(N)$.

The most interesting question is whether an analogous study can be performed for black hole solutions in ${\rm hs}[\lambda]$ Vasiliev theory which is related by the proposal of Gaberdiel and Gopakumar to a 't~Hooft large-$N$ limit of the ${\cal W}_N$ minimal models. Higher spin black hole solutions in this theory, constructed perturbatively in the chemical potential \cite{krausperl}, do alter the asymptotics in the same way as  in ${\rm SL}(N, \mathbb R)$ Chern-Simons theories. It would definitely be interesting to understand if these imply an RG flow between a ${\cal W}_N$ minimal model and  another (unitary)  CFT. Non-principal embeddings of certain kinds (e.g. the next-to-principal embedding) have been argued to have semiclassical limits which are unitary \cite{Afshar:2012hc}. In any case, it appears that ${\rm SL}(N)$ Chern-Simons theories in the semiclassical limit (fixed $N$, large $k$) yield results related by analytic continuation to the ${\rm hs}[\lambda]$ theory or the unitary ${\cal W}_N$ models in the 't~Hooft limit \cite{Gaberdiel:2012ku, Hijano:2013fja}. Therefore, it is plausible that a similar picture is realized for RG flows within this framework. 

The analysis of the thermodynamics of multiple branches of black hole solutions in the large-$N$ limit is another fascinating question for the future. We have already seen that the number of solutions to the holonomy equations increases rapidly with $N$. 
Since the BTZ-branch of solutions disappears for high temperatures
(as a real solution to the holonomy conditions) this is accompanied by putative discontinuous phase transitions \cite{David:2012iu}. The discretuum of solutions at large-$N$ may well be responsible for smoothing out and eliminating such discontinuities as seen at zero chemical potential 
\cite{Banerjee:2012aj}. Conical defect states \cite{conical, Campoleoni:2013lma} carrying higher spin charge, are also likely to play an important role in this limit.

There are very close and interesting parallels between the RG flows 
associated to higher spin black holes as we have described  and the theory of generalized KdV hierarchies \cite{DeGroot:1991ca, Burroughs:1991bd}. The latter are naturally associated to the
affine generalisation of $\mg=\mf{sl}(N)$ realized as a loop algebra $\widehat{\mg}=\mf{sl}(N)\otimes{\mathbb C}[\xi,\xi^{-1}]$.\footnote{We denote the spectral parameter as $\xi$ to avoid confusion with the spacetime variable $z$.} The connection with the RG flows is that the Chern-Simons equation-of-motion is a flatness condition for the gauge connection $(a_z,a_{\bar z})$ which can be interpreted as the Lax equation of the hierarchy associated to the pair of  ``flows" $(z,\bar z)$. In the context of the integrable hierarchies, a central role is played by a particular Heisenberg subalgebra of $\widehat{\mg}$. These are maximal abelian subalgebras of the affine algebra which are classified by conjugacy classes of $\mg$. In the present context, we need to take the Heisenberg subalgebra associated to the Coxeter element of the Weyl group of $\mf{sl}(N)$ which in the $N$-dimensional representation take the form
\EQ{
\Lambda^{(s,n)}=\xi^n\big(W^{(s)}_{s-1}+\xi Y_{s-1-N}\big)\ ,
}
where $Y_{s-1-N}$ is a particular element with $[L_0,Y_{s-1-N}]=(s-1-N)Y_{s-1-N}$ determined by the condition that
$[\Lambda^{(s,n)},\Lambda^{(s',n')}]=0$.
The basic Lax operator of the hierarchy is 
\EQ{
L=\partial_z+q+\Lambda^{(1,0)}\ .
}
The flows of the hierarchy are then given by
\EQ{
\frac{\partial L}{\partial t^{(s,n)}}=[L,A^{(s,n)}]\ ,
}
The definition of $q$ and $A^{(s,n)}$ above is described fully in \cite{DeGroot:1991ca}, however, in the present context we can identify $q+\Lambda^{(1,0)}=a_z$ (with $\xi=0$). 
The RG flow when a given $\mu_s$ is turned on is then associated to a particular flow of the hierarchy with $\bar z=t^{(s,0)}$ and $A^{(s,0)}=a_{\bar z}$ (again with $\xi=0$). The fact that the equations are integrable means that there are an infinite number of conserved quantities which in turn means that the Ward identities of the $2d$ QFT that interpolates between the UV and IR CFTs can be written as an infinite set of conservation equations. This implies that the non-relativistic interpolating QFT is integrable. Another important property of the integrable hierarchy is that the equations can be written in Hamiltonian form for each element of the Heisenberg subalgebra \cite{Burroughs:1991bd}. The corresponding Poisson structure associated to the flow $t^{(s,0)}$ is then naturally identified with the generalized ${\cal W}$-algebra for the non-principal embedding with $\hat L_1=W^{(s)}_{s-1}$.

\acknowledgments

We would like to thank Justin David for several enjoyable discussions and for collaboration during early stages of the project. The authors were supported in part by STFC grant ST/G000506/1.
\newpage
\appendix
\section{Conventions}
\label{conventions}
We follow the conventions adopted in \cite{conical} for defining and obtaining matrix representations of $\mf{sl}(N,\mathbb R)$ elements. After taking $\{L_0,L_{\pm1}\}$ to be the $N$-dimensional representations of $\mf{sl}(2)$, the explicit form for the other generators can be deduced using
\bea 
W_m^{(s)}\,=\,(-1)^{s-m-1}\frac{(s+m-1)!}{(2s-2)!}\,
\underbrace{[L_{-1},[L_{-1},\ldots[L_{-1}}_{s-m-1 \,{\rm terms}},L_1^{s-1}]\ldots]]
\eea
\section{Decomposition of $\mf{sl}(4)$ Algebra}
\label{appa}
The $\mf {sl}(4)$ generators branch into
irreducible representations of $\mf{sl}(2)$ as,
\be
{\bf 15} \to \underline{\bf 3}\,+\,4\cdot\underline{\bf 2}\,+\,4\cdot \underline{\bf 1}\,.
\ee
and the algebra naturally decomposes as 
$\mf {sl}(4) \to \mf{sl}(2)\oplus \mf{sl}(2)'\oplus \mf{u}(1)$. 
Explicitly, under this branching, the 15 generators of $\mf{sl}(4)$ are:
\bea
&&\mf{sl}(2):\qquad\hat L_0\,=\, \tfrac{1}{3}\,L_0 + \tfrac{1}{3}
\left(\tfrac{1}{2} W^{(4)}_0-\tfrac{1}{10}L_0 \right)\,,\qquad \hat L_{\pm 1}\,=\,
-\tfrac{1}{6}\,W^{(4)}_{\pm 3}\,,\label{gen}\\\nonumber\\\nonumber
&&\mf{sl}(2)':\qquad { J}_0\,=\,\tfrac{1}{10}L_0 -\tfrac{1}{2}W^{(4)}_0\,,\qquad
J^{\pm}\,=\,\tfrac{1}{5}L_{\pm 1} -\tfrac{1}{2}W^{(4)}_{\pm1}\,,
\\\nonumber\\\nonumber
&&\mf{u}(1):\qquad J\,=\, \tfrac{1}{2}W^{(3)}_0\,,\\\nonumber\\\nonumber
&& G_{\tiny{{1}/{2}}}^{+\,\pm}\,=\,\tfrac{1}{2\sqrt 3}\,\left(-W^{(4)}_{2}\pm\tfrac{1}{2}W^{(3)}_{2}\right)\,,\qquad G_{-1/2}^{-\,\pm}\,=\,\tfrac{1}{2\sqrt 3}\,\left(W^{(4)}_{-2}\pm\tfrac{1}{2}W^{(3)}_{-2}\right)\,,\\\nonumber\\\nonumber
&& G_{\pm1/2}^{\mp\, \pm}\,=\,\pm\,\tfrac{1}{2\sqrt 3}\left(W^{(3)}_{\pm 1} - W^{(4)}_{\pm 1} -\tfrac{3}{5}L_{\pm 1}\right)\,,
\,\,\, G_{\mp 1/2}^{\pm\,\pm}\,=\,\pm\,\tfrac{1}{2\sqrt 3}\left(
\tfrac{3}{5}L_{\mp 1}+W^{(4)}_{\mp 1}+W^{(3)}_{\mp 1}\right)\,.
\eea
The generators $G_m^{ab}$ are labelled by their weights $m=\pm \tfrac{1}{2}$, $a=\pm$, $b=\pm$ with respect  to  
$\hat L_0$, $J$ and $J^{0}$, respectively. The algebra of these generators is the global part of the ${\cal W}^{(2,1,1)}_4$ algebra. 

\section{Variations of SL$(\boldsymbol 4,\mathbb R)$ Currents}
\label{appvar}

We list below the transformation laws for the currents under infinitesimal $\text{SL}(4,\mathbb R)$ transformations that preserve the form of the asymptotically AdS background in the $(2,1,1)$ embedding. In all equations below we  drop the subscript on $\chi_+^{ab}$. The non-trivial gauge transformations are:\\
{\underline{Variations\, of\, $\hat {\cal L}$}:}
\bea
&&\delta_\varepsilon \hat{\cal L}\,=\,-\tfrac{1}{2}\,\varepsilon_1'''+2 \hat{\cal L}\,\varepsilon_1'
+\varepsilon_1\,\hat{\cal L}'\,,\label{var1}\\\nonumber\\\nonumber
&&\delta_{\chi^{++}}\hat{\cal L}\,=\,-\tfrac{3}{2}\chi^{++\,'}{\cal G}_{--}+\left( -\tfrac{1}{2}{\cal G}_{--}'-{\cal G}_{--}\left(\EuScript J-\tfrac{1}{2}{\cal J}_0\right)\,-{\cal G}_{+-}\,{\cal J}_-\right)\,\chi^{++},\\\nonumber\\\nonumber
&&\delta_{\chi^{-+}}\hat{\cal L}\,=\,\tfrac{3}{2}\chi^{-+\,'}{\cal G}_{+-}+
\left( \tfrac{1}{2}{\cal G}_{+-}'+{\cal G}_{+-}\left(\EuScript J+\tfrac{1}{2}{\cal J}_0\right)\,-{\cal G}_{--}\,{\cal J}_+\right)\,\chi^{-+}\,,\\\nonumber\\\nonumber
&&\delta_{\chi^{+-}}\hat{\cal L}\,=\,-\tfrac{3}{2}\chi^{+-\,'}{\cal G}_{+-}+\left(-\tfrac{1}{2}{\cal G}_{-+}'+{\cal G}_{-+}\left(\EuScript J+\tfrac{1}{2}{\cal J}_0\right)\,-{\cal G}_{++}\,{\cal J}_-\right)\,\chi^{+-},\\\nonumber\\\nonumber
&&\delta_{\chi^{--}}\hat{\cal L}\,=\,\tfrac{3}{2}\chi^{--\,'}\,{\cal G}_{++} +\left( \tfrac{1}{2}{\cal G}_{++}'-{\cal G}_{++}\left(\EuScript J-\tfrac{1}{2}{\cal J}_0\right)\,-{\cal G}_{-+}\,{\cal J}_+\right)\,\chi^{--}\,.
\eea
{ \underline{Variations\, of\, \U(1)\, current}:}
\bea
&&\delta_{\gamma}\EuScript J\,=\,\gamma'\,\label{var2}
\label{var2}\\\nonumber\\\nonumber
&&\delta_{\chi^{++}}\EuScript J\,=\,{\cal G}_{--}\,\chi^{++}\,,\qquad
\delta_{\chi^{--}}\EuScript J\,=\,{\cal G}_{++}\,\chi^{--}\,,\\\nonumber\\\nonumber
&&\delta_{\chi^{+-}}\EuScript J\,=\,-{\cal G}_{-+}\,\chi^{+-}\,,\qquad
\delta_{\chi^{-+}}\EuScript J\,=\,-{\cal G}_{+-}\,\chi^{-+}\,.
\eea
{\underline{Variations\, of\, {${\rm SL}(2)^\prime$}\, currents}}:
\bea
&&\delta_{\eta_0}{\cal J}_0\,=\,\eta_0'\,,\quad \delta_{\eta_-}{\cal J}_0\,=\,2{\cal J}_+\,\eta_-\,,\quad \delta_{\eta_+}{\cal J}_0\,=\,-2{\cal J}_-\,\eta_+\,,
\quad\delta_{\chi^{--}}{\cal J}_0\,=\,-{\cal G}_{++}\,\chi^{--}\,,\nonumber\\\nonumber\\\label{var3}
&&
\delta_{\chi^{++}}{\cal J}_0\,=\,-{\cal G}_{--}\,\chi^{++}\,,\quad
\delta_{\chi^{+-}}{\cal J}_0\,=\,-{\cal G}_{-+}\,\chi^{+-}\,,\quad
\delta_{\chi^{-+}}{\cal J}_0\,=\,-{\cal G}_{+-}\,\chi^{-+}\,.\\\nonumber\\\nonumber
&&\delta_{\eta_+}{\cal J}_+\,=\,\eta_+'-2{\cal J}_0\,\eta_+\,;\,\,
\delta_{\eta_0}{\cal J}_+\,=\,{\cal J}_+\,\eta_0\,;\,\,
\delta_{\chi^{+-}}{\cal J}_+\,=\,-{\cal G}_{++}\chi^{+-}\,;\,\,
\delta_{\chi^{++}}{\cal J}_+\,=\,-{\cal G}_{+-}\chi^{++}\,.\nonumber
\\\nonumber\\\nonumber
&&\delta_{\eta_-}{\cal J}_-\,=\,\eta_-'+{\cal J}_0\,\eta_-\,;\,\,
\delta_{\eta_0}{\cal J}_-\,=\,-{\cal J}_-\,\eta_0\,;\,\,
\delta_{\chi^{-+}}{\cal J}_-\,=\,-{\cal G}_{--}\chi^{-+}\,;\,\,
\delta_{\chi^{--}}{\cal J}_-\,=\,-{\cal G}_{-+}\chi^{--}\,.\nonumber
\eea
{\underline{Variations of spin-$\tfrac{3}{2}$ currents}:}
\bea
&&\delta_\varepsilon{\cal G}_{\pm+}\,=\,\tfrac{3}{2}\,{\cal G}_{\pm+}\,\,\varepsilon_1'\,+\,
\left({\cal G}_{\pm+}'\,+\,{\cal G}_{\pm+}\,\EuScript J\,\mp\,
\tfrac{1}{2}{\cal G}_{\pm+}\,{\cal J}_0\,
\pm\,{\cal G}_{\mp+}\,{\cal J}_\pm\right)\,\varepsilon_1\,,
\label{var4}\\\nonumber\\\nonumber
&&\delta_\varepsilon{\cal G}_{\pm-}\,=\,\tfrac{3}{2}\,{\cal G}_{\pm-}\,\varepsilon_1'\,+
\,\left({\cal G}_{\pm-}'\,-\,{\cal G}_{\pm-}\,\EuScript J\,\mp\,\tfrac{1}{2}{\cal G}_{\pm-}\,{\cal J}_0\,
\mp\,{\cal G}_{\mp-}\,{\cal J}_\pm\right)\,\varepsilon_1\,,
\eea
\bea
&&\delta_{\chi^{++}}{\cal G}_{++}\,=\,-\chi^{++\,''}+\chi^{++\,'} \,({\cal J}_0-2\EuScript J)
+\chi^{++}\,\left(\tfrac{1}{2}{\cal J}_0'-\EuScript J'+{\cal J}_+\,{\cal J}_- -\tfrac{1}{4}{\cal J}_0^2+\EuScript J\,{\cal J}_0-\EuScript J^2 + \hat{\cal L}\right)\nonumber\\\nonumber\\
&&\delta_{\chi^{-+}}{\cal G}_{++}\,=\,-\,2{\cal J}_+\chi^{-+\,'}\,-\,(2\EuScript J\,{\cal J}_++{\cal J}_+')\chi^{-+}
\\\nonumber\\\nonumber
&&\delta_{\gamma}{\cal G}_{++}\,=\,-\,{\cal G}_{++}\,\gamma\,\qquad
\delta_{\eta_0}{\cal G}_{++}\,=\,\tfrac{1}{2}{\cal G}_{++}\,\eta_0\qquad
\delta_{\eta_+}{\cal G}_{++}\,=\,-\,{\cal G}_{-+}\,\eta_+
\\\nonumber\\\nonumber
&&\delta_{\chi^{-+}}{\cal G}_{-+}\,=\,-\chi^{-+\,''}-\chi^{-+\,'} \,({\cal J}_0+2\EuScript J)
+\chi^{-+}\,\left(-\tfrac{1}{2}{\cal J}_0'-\EuScript J'+{\cal J}_+\,{\cal J}_- -\tfrac{1}{4}{\cal J}_0^2-\EuScript J\,{\cal J}_0-\EuScript J^2 +\hat{\cal L}\right)\\\nonumber\\\nonumber
&&\delta_{\chi^{++}}{\cal G}_{-+}\,=\,2{\cal J}_-\chi^{++\,'}\,+\,(2\EuScript J\,{\cal J}_-+{\cal J}_-')\chi^{++}
\\\nonumber\\\nonumber
&&\delta_{\gamma}{\cal G}_{-+}\,=\,-\,{\cal G}_{-+}\,\gamma\,\qquad
\delta_{\eta_0}{\cal G}_{-+}\,=\,-\tfrac{1}{2}{\cal G}_{-+}\,\eta_0\qquad
\delta_{\eta_-}{\cal G}_{-+}\,=\,{\cal G}_{++}\,\eta_-
\\\nonumber\\\nonumber
&&\delta_{\chi^{--}}{\cal G}_{--}\,=\,-\chi^{--\,''}-\chi^{--\,'} \,({\cal J}_0-2\EuScript J)
+\chi^{--}\,\left(-\tfrac{1}{2}{\cal J}_0'+\EuScript J'+{\cal J}_+\,{\cal J}_- -\tfrac{1}{4}{\cal J}_0^2+\EuScript J\,{\cal J}_0-\EuScript J^2 +\hat{\cal L}\right)\\\nonumber\\\nonumber
&&\delta_{\chi^{+-}}{\cal G}_{--}\,=\,2{\cal J}_-\chi^{+-\,'}\,+\,(-2\EuScript J\,{\cal J}_-+{\cal J}_-')\chi^{+-}
\\\nonumber\\\nonumber
&&\delta_{\gamma}{\cal G}_{--}\,=\,\,{\cal G}_{--}\,\gamma\,\qquad
\delta_{\eta_0}{\cal G}_{--}\,=\,-\tfrac{1}{2}{\cal G}_{--}\,\eta_0\qquad
\delta_{\eta_-}{\cal G}_{--}\,=\,{\cal G}_{+-}\,\eta_-
\\\nonumber\\\nonumber
&&\delta_{\chi^{+-}}{\cal G}_{+-}\,=\,-\chi^{+-\,''}+\chi^{+-\,'} \,({\cal J}_0+2\EuScript J)
+\chi^{+-}\,\left(\tfrac{1}{2}{\cal J}_0'+\EuScript J'+{\cal J}_+\,{\cal J}_- -\tfrac{1}{4}{\cal J}_0^2-\EuScript J\,{\cal J}_0-\EuScript J^2 +\hat{\cal L}\right)\\\nonumber\\\nonumber
&&\delta_{\chi^{--}}{\cal G}_{+-}\,=\,-2{\cal J}_+\chi^{+-\,'}\,+\,(2\EuScript J\,{\cal J}_+-{\cal J}_+')\chi^{--}
\\\nonumber\\\nonumber
&&\delta_{\gamma}{\cal G}_{+-}\,=\,\,{\cal G}_{+-}\,\gamma\,\qquad
\delta_{\eta_0}{\cal G}_{+-}\,=\,\tfrac{1}{2}{\cal G}_{+-}\,\eta_0\qquad
\delta_{\eta_+}{\cal G}_{+-}\,=\,-{\cal G}_{--}\,\eta_+\,.
\eea
\subsection{Poisson brackets for the ${\cal W}_4^{(2,1,1)}$ algebra} 
Using the gauge variations of the currents, we arrive at the following set of Poisson brackets which determine the ${\cal W}$-algebra. In order to express our results in terms of appropriately normalized currents, we have performed
the rescalings
\be
\hat{\cal L} \to -\tfrac{2\pi}{k}\hat{\cal L}\,,\quad {\cal G}_{ab}\to
-\tfrac{2\pi}{k}{\cal G}_{ab}\,,\quad{\cal J}_\pm\to-\tfrac{2\pi}{k}{\cal J}_\pm\,,\quad {\cal J}_0\to-\tfrac{4\pi}{k}{\cal J}_0\,,\quad \EuScript J \to -\tfrac{2\pi}{k}\EuScript J
\label{rescale}
\ee
\bea
&&\{\hat{\cal L}(z),\hat{\cal L}(z')\}\,=\,- 2\hat{\cal L}(z)\,\delta'(z-z')\,-\,\hat{\cal L}'\,\delta(z-z') \,-\,
\tfrac{k}{4\pi}\delta'''(z-z')\,,\label{PB}\\\nonumber\\\nonumber
&&\{\hat{\cal L}(z),{\cal G}_{\pm+}(z')\}\,=\,-\tfrac{3}{2}\,
\delta'(z-z')\,{\cal G}_{\pm+}(z)\, -\, \tfrac{1}{2}\,\delta(z-z')\,
{\cal G}_{\pm+}'(z)\\\nonumber&&\hspace{2.4in}-\tfrac{2\pi}{k}\,
\left({\cal G}_{\pm+}\,\EuScript J\,\mp\,{\cal G}_{\pm +}\,{\cal J}_0\,\pm \,{\cal G}_{\mp+}\,{\cal J}_\pm\right)\,\delta(z-z')\,,
\eea
\bea
&&\{\hat{\cal L}(z),{\cal G}_{\pm-}(z')\}\,=\,-\tfrac{3}{2}\,
\delta'(z-z')\,{\cal G}_{\pm -}(z) - \tfrac{1}{2}\,\delta(z-z')
{\cal G}_{\pm-}'(z)\nonumber\\\nonumber&&\hspace{2.5in}-\tfrac{2\pi}{k}\,
\left(-{\cal G}_{\pm-}\,\EuScript J\,\mp\,{\cal G}_{\pm -}\,{\cal J}_0
\mp \,{\cal G}_{\mp-}\,{\cal J}_\pm\right)\,\delta(z-z')\,,
\\\nonumber\\\nonumber
&&\{\hat{\cal L}(z),\EuScript J(z')\}\,=\,0\,,\qquad
\{\hat{\cal L}(z),{\cal J}_{0,\pm}(z')\}\,=\,0\,,\qquad\{\EuScript J(z),\EuScript J(z')\}\,=\,\tfrac{k}{2\pi}\,\delta'(z-z')\,,\\\nonumber\\\nonumber
&&\{{\cal J}_0(z),{\cal J}_0(z')\}\,=\,\tfrac{k}{4\pi}\delta'(z-z')\,,
\qquad\{{\cal J}_0(z),{\cal J}_\pm(z')\}\,=\,\mp\,\tfrac{k}{2\pi}\,{\cal J}_\pm\,\delta(z-z')\,,
\\\nonumber\\\nonumber
&&\{{\cal J}_+(z),{\cal J}_-(z')\}\,=\,-\tfrac{k}{2\pi}\delta'(z-z')\,-\,2{\cal J}_0\,\delta(z-z')\,,
\\\nonumber\\\nonumber
&&\{{\cal G}_{\pm\mp}(z),{\cal G}_{\pm\pm}(z')\}\,=\, 2\,{\cal J}_\pm(z)\,\delta'(z-z')\,+
{\cal J}_\pm'\,\delta(z-z')\,\pm\,\tfrac{4\pi}{k}\EuScript J\,{\cal J}_\pm\,\delta(z-z')\,,\\\nonumber\\\nonumber
&&\{{\cal G}_{-\mp}(z),{\cal G}_{+\pm}(z')\}\,=\,-\tfrac{k}{2\pi}\delta''(z-z')\,+\,\delta'(z-z')\,(2{\cal J}_0(z)\mp2\EuScript J(z))\,-\,\delta(z-z')\,{\cal L}\\\nonumber
&&\qquad\qquad\qquad +\,\delta(z-z')\,({\cal J}_0'(z)\,\mp\,\EuScript J'(z))
+\,\tfrac{2\pi}{k}\delta(z-z')\,({\cal J}_+\,{\cal J}_-\,-\,{\cal J}_0^2\,\pm\,2j\,{\cal J}_0\,-\,j^2\,)\,,\\\nonumber\\\nonumber
&&\{\EuScript J(z),{\cal G}_{a \pm}(z')\}\,=\,\mp\delta(z-z')\,{\cal G}_{a,\pm}\,,\qquad
\{{\cal J}_0(z),{\cal G}_{\pm,a}(z')\}\,=\,\pm\frac{1}{2}\delta(z-z')\,{\cal G}_{\pm,a}\,,\\\nonumber\\\nonumber
&&\{{\cal J}_\mp(z),{\cal G}_{\pm,a}(z')\}\,=\,
\pm\, {\cal G}_{\mp, a}\,\delta(z-z')\,.
\eea
These Poisson brackets define the ${\cal W}_4^{(2,1,1)}$ algebra. A final adjustment is necessary in order to ensure that the currents have standard tensor-like transformation properties. This is easily achieved by a shift of $\hat{\cal L}$ which yields the correct stress tensor, \be
\hat{\cal L}\,\to\, T\,=\,\hat{\cal L}-\tfrac{2\pi}{k}\left(\tfrac{1}{2}\EuScript J^2+
{\cal J}_0^2 -{\cal J}_+\,{\cal J}_-\right)\,.
\ee
We then obtain the following Poisson brackets involving the stress tensor $T$,
\bea
&&\{T(z),{\cal G}_{ab}(z')\}\,=\,
-\tfrac{3}{2}\,\delta'(z-z')\,{\cal G}_{ab}(z)\,-\,\tfrac{1}{2}
\,\delta(z-z')\,{\cal G}'_{ab}(z)\,,\\\nonumber\\\nonumber
&&\{T(z), \EuScript J(z')\}\,=\,-\delta'(z-z')
\,\EuScript J(z)\,,\qquad \{T(z), {\cal J}_{A}(z')\}\,=\,-\delta'(z-z')
\,{\cal J}_A(z)\,,\\\nonumber\\\nonumber
&&\{{\cal G}_{-\mp}(z),{\cal G}_{+\pm}(z')\}\,=\,\delta''(z-z')\,+\,\delta'(z-z')\,(2{\cal J}_0(z)\mp2\EuScript J(z))\,-\,\delta(z-z')\,T(z)\\\nonumber
&&\,\,\,+\,\delta(z-z')\,({\cal J}_0'(z)\,\mp\,\EuScript J'(z))
+\,\tfrac{2\pi}{k}\delta(z-z')\,(2{\cal J}_+\,{\cal J}_-\,-\,2{\cal J}_0^2\,\pm\,j\,{\cal J}_0\,-\,\tfrac{3}{2}j^2\,)\,.
\eea
The Poisson brackets above define the ${\cal W}$-algebra associated to the non-principally embedded Chern-Simons connection

\section{Dictionary Between ${\cal W}_4^{(2,1,1)}$ Currents and Black Hole Connection}
\label{dict}

The gauge transformation $\Omega$ (eq.~\eqref{gauge}) on the generalized ${\rm SL}(4,\mathbb R)$ black hole connection \eqref{uvsl4}, turns it into the general form shown in eq.~\eqref{asympz} allowing us to read off the parameters of the black hole background in terms of ${\cal W}$-algebra currents,
\bea
&&v_0\,=\,-\tfrac{1}{2}\,\EuScript J\,,\quad v_{-1}\,=\,\tfrac{1}{2\sqrt 3}\,\left({\cal G}_{+-} - {\cal G}_{++}\right)\,,\quad v_{-2}\,=\,\tfrac{1}{12}\,\left(\sqrt3\,{\cal G}_{--}-\sqrt3\,{\cal G}_{-+}
-2\EuScript J\,{\cal J}_+\right)\nonumber\\\nonumber\\\nonumber
&& w_1\,=\,\tfrac{5}{6}\,{\cal J}_+\,,\quad w_0\,=\,\tfrac{5}{9}\,{\cal J}_0\,,
\quad w_{-1}\,=\,\tfrac{1}{2}\,{\cal J}_- -\tfrac{1}{6}{\cal J}_+^2-\tfrac{1}{2\sqrt 3}\left({\cal G}_{+-}+{\cal G}_{++}\right)\,,\\\nonumber\\\nonumber
&&w_{-2}\,=\, \tfrac{1}{3}\bar\partial{\cal J}_+-\tfrac{1}{9}\,{\cal J}_0{\cal J}_+-\tfrac{1}{2\sqrt 3}\left({\cal G}_{--}+{\cal G}_{-+}\right)\,,\\\nonumber\\\nonumber
&&w_{-3}\,=\,-\tfrac{1}{6}\hat{\cal L}+\tfrac{1}{36}\bar\partial {\cal J}_0+\tfrac{1}{18}\,{\cal J}_+^3+\tfrac{1}{216}\,{\cal J}_0^2+\tfrac{1}{6\sqrt 3}\left({\cal G}_{+-}+{\cal G}_{++}\right)\,{\cal J}_+\,,\\\nonumber\\
&&u_{-1}\,=\,-\tfrac{1}{5}\,{\cal J}_--\tfrac{4}{15}\,{\cal J}_+^2-\tfrac{\sqrt 3}{10}\,\left({\cal G}_{+-}+{\cal G}_{++}\right)\,.\label{dictionary}
\eea
A further rescaling as indicated in eq.~\eqref{rescale} is necessary in order to express the result in terms of correctly normalized currents.

\end{document}